\documentclass[sigconf]{acmart}

\AtBeginDocument{%
  }

\setcopyright{acmcopyright}
\copyrightyear{2024}
\acmYear{2024}
\acmDOI{3597503.3639164}

\acmConference[ICSE 2024]{46th International Conference on Software Engineering}{April 2024}{Lisbon, Portugal}
\acmPrice{15.00}
\acmISBN{979-8-4007-0217-4/24/04}

\usepackage{xcolor}
\usepackage{graphicx}
\usepackage{textcomp}
\usepackage{subfigure}
\usepackage{hyperref}
\usepackage{array,booktabs}
\usepackage{multirow}
\usepackage{balance}

\usepackage{tikz}

\usepackage[normalem]{ulem}

\usepackage{tcolorbox}

\newcommand{\PreserveBackslash}[1]{\let\temp=\\#1\let\\=\temp}
\newcolumntype{C}[1]{>{\PreserveBackslash\centering}p{#1}}
\newcolumntype{R}[1]{>{\PreserveBackslash\raggedleft}p{#1}}
\newcolumntype{L}[1]{>{\PreserveBackslash\raggedright}p{#1}} 
\begin{document}


%\title{TRIAD: Automated Traceability Recovery based on Biterm-enhanced Deduction of Transitive Links from Multiple Artifacts}

\title{TRIAD: Automated Traceability Recovery based on Biterm-enhanced Deduction of Transitive Links among Artifacts}

\author{Hui Gao} 
\affiliation{%
  \institution{State Key Lab of Novel Software}
  \institution{ Technology, Nanjing University}
  \city{Nanjing} 
  \country{China}}
\email{ghalexcs@gmail.com}

\author{Hongyu Kuang} 
\affiliation{%
  \institution{State Key Lab of Novel Software}
  \institution{ Technology, Nanjing University}
  \city{Nanjing} 
  \country{China}}
\email{khy@nju.edu.cn}
\authornote{Corresponding author}

\author{Wesley K. G. Assunção} 
\affiliation{%
  \institution{CSC, North Carolina State University}
  \city{Raleigh} 
  \country{USA}}
\email{wguezas@ncsu.edu}

\author{Christoph Mayr-Dorn} 
\affiliation{%
  \institution{ISSE, Johannes Kepler University}
  \city{Linz} 
  \country{Austria}}
\email{christoph.mayr-dorn@jku.at}

\author{Guoping Rong} 
\affiliation{%
  \institution{State Key Lab of Novel Software}
  \institution{ Technology, Nanjing University}
  \city{Nanjing} 
  \country{China}}
\email{ronggp@nju.edu.cn}

\author{He Zhang} 
\affiliation{%
  \institution{State Key Lab of Novel Software}
  \institution{ Technology, Nanjing University}
  \city{Nanjing} 
  \country{China}}
\email{hezhang@nju.edu.cn}

\author{Xiaoxing Ma} 
\affiliation{%
  \institution{State Key Lab of Novel Software}
  \institution{ Technology, Nanjing University}
  \city{Nanjing} 
  \country{China}}
\email{xxm@nju.edu.cn}

\author{Alexander Egyed} 
\affiliation{%
  \institution{ISSE, Johannes Kepler University}
  \city{Linz} 
  \country{Austria}}
\email{alexander.egyed@jku.at}

\begin{abstract}
%Context: 
Traceability allows stakeholders to extract and comprehend the trace links among software artifacts introduced across the software life cycle, to provide significant support for software engineering tasks.
Despite its proven benefits, software traceability is challenging to recover and maintain manually.
Hence, plenty of approaches for automated traceability have been proposed. 
Most rely on textual similarities among software artifacts, such as those based on Information Retrieval (IR).
%Problem
However, artifacts in different abstraction levels usually have different textual descriptions, which can greatly hinder the performance of IR-based approaches (e.g., a requirement in natural language may have a small textual similarity to a Java class). 
%Solutions: 
In this work, we leverage the consensual biterms and transitive relationships (i.e., inner- and outer-transitive links) based on intermediate artifacts to improve IR-based traceability recovery.
We first extract and filter biterms from all source, intermediate, and target artifacts.
We then use the consensual biterms from the intermediate artifacts to enrich the texts of both source and target artifacts, and finally deduce outer and inner-transitive links to adjust text similarities between source and target artifacts.
%Evaluation: 
We conducted a comprehensive empirical evaluation based on five systems widely used in other literature to show that our approach can outperform four state-of-the-art approaches in Average Precision over 15\% and Mean Average Precision over 10\% on average.
\end{abstract}




\maketitle

%\balance

\section{Introduction \& Motivation}
\label{sec:introduction}
%Context:
Multiple software artifacts in different levels of abstraction are introduced during the development process, including (1) high-level requirements, use cases, and specifications; and (2) low-level artifacts such as source code and tests~\cite{Grundy1998, Ivkovic/ICSM04}. 
Accordingly, software traceability is defined as ``the ability to interrelate any uniquely identifiable software engineering artifact to any other, maintain required links over time, and use the resulting network to answer questions of both the software product and its development process''~\cite{CoEST}.
It allows stakeholders to comprehend the system functionalities behind introduced software artifacts. 
Recent work has shown that traceability is paramount in the context of collaborative development, helping developers to keep all artifacts synchronized and consistent across a myriad of tools and domains~\cite{Cosmina2022}.
When correctly recovered and maintained, these trace links can not only help improve the software quality\cite{DBLP:journals/tse/RempelM17} and the efficiency of software maintenance\cite{journals/ese/MaderE15}, but also provide support for many critical software engineering tasks such as safety-assurance\cite{conf/icse/Mayr-DornVBKCEM21}, change impact analysis\cite{journals/tse/FalessiRGC20}, and bug localization\cite{conf/msr/0002LM08}.
In practice, traceability is also widely used to demonstrate the safe running of systems\cite{DBLP:journals/infsof/NejatiSFBC12, DBLP:journals/software/MaderJZC13}, and to help ensure system security\cite{DBLP:conf/icse/NhlabatsiYZTKBK15, DBLP:conf/icse/MoranPBMPSJ20}.
Unfortunately, despite its importance, manually creating and maintaining trace links is time-consuming and error-prone \cite{DBLP:journals/tse/RameshJ01, DBLP:conf/re/EgyedGG10} due to the semantic gap between different artifacts \cite{conf/icse/BiggerstaffMW93} (such as requirements written in natural language and code written in programming language).

Therefore, researchers have investigated the use of automated tracing techniques based on information retrieval (IR)\cite{DBLP:conf/iwpc/KuangG0M0ME19, DBLP:journals/ese/GaoKMHLME22}, machine learning (ML)\cite{conf/icsm/MillsEH18, conf/icsm/MillsEBKCH19}, and deep learning techniques (DL)\cite{DBLP:conf/icse/0004CC17, conf/icse/LinLZ0C21}. 
These automated approaches generally rely on analyzing the textual similarities or relationships from different artifact texts to derive the likelihood of the actual traces. 
Unfortunately, different types of software artifacts typically have different abstraction levels due to being produced in different stages of the software life-cycle \cite{ADeLuciaSST2012}.
Specifically, the early stages deal with more user-and-problem-centric concepts (e.g., system specifications and requirements), while later stages emphasize solution and implementation details (e.g., code and test)\cite{journals/software/ChikofskyC90}. 
Spanning life-cycle phases involves a detailed transition from higher abstraction levels in early stages to lower abstraction levels in later stages, thus causing the \textit{abstraction gap problem} among different artifacts at different abstraction levels, i.e., one software concept is often described in summarized form in high-level abstracted artifacts, while the description of the same concept is available in more detailed text in low-level abstracted artifacts.
Furthermore, different kinds of software artifacts usually use different terms to express the same concept\cite{conf/icse/BiggerstaffMW93}, which is also known as the \textit{vocabulary mismatch problem}.
These two situations greatly hinder the performance of automated tracing techniques.

To address this issue, researchers proposed diverse enhancing strategies from different perspectives for automated traceability, including text enrichment\cite{DBLP:conf/re/Cleland-HuangSDZ05, DBLP:conf/iwpc/DiazBMOTL13}, advanced lexical analysis\cite{DBLP:conf/icsm/GethersOPL11, Panichella6606598}, incorporating execution tracing by running systems\cite{DBLP:journals/tse/PoshyvanykGMAR07, DBLP:journals/ese/DitRP13}, analyzing code structural information\cite{DBLP:conf/icse/McMillanPR09, DBLP:conf/wcre/KuangNHRLEM17}, advanced data pre-processing and models for ML and DL\cite{conf/icse/LinLZ0C21, DBLP:conf/sigsoft/0001ZLWK22}, and utilizing user verification on candidate links\cite{DBLP:journals/tse/HayesDS06, DBLP:conf/icsm/LuciaOS06, DBLP:conf/csmr/PanichellaMMPOPL13, DBLP:conf/icse/PanichellaLZ15, DBLP:journals/ese/GaoKMHLME22}. 
However, the effectiveness of these approaches still depends on the quantity and quality of artifact texts. 
Furthermore, these approaches focused on establishing trace links by solely considering the two trace endpoint artifacts (usually requirements and code that have perhaps the largest abstraction gap), ignoring other available intermediate artifacts that are simultaneously created and maintained during development, such as various design artifacts and test cases. 

A different body of work focused on introducing other types of artifacts in automated traceability to bridge the abstraction gap. 
Moran et al.\cite{DBLP:conf/icse/MoranPBMPSJ20} proposed to recover requirements-to-code traces by incorporating user feedback with transitive links that are additionally extracted between requirements and test cases.   
Rodriguez et al.\cite{conf/icsm/RodriguezCF21} explored 84 kinds of combinations for eliciting transitive links from intermediate artifacts to recover source-to-target traces. 
However, it is also reported that the optimized technique across all datasets is hard to locate. The users have to fine-tune the combination strategies for their systems to reach satisfying performance.
Meanwhile, Gao et al.\cite{Hui2022ASE} proposed to extract consensual biterms from both requirements and code to improve IR-based traceability recovery. They argue that the extracted biterms indicate part of the same system functionalities shared between requirements and code, thus helping to bridge their semantic gaps.
However, their biterm extraction does not involve available intermediate artifacts that can be vital to further solve the abstraction gap problem.

To break through the previously discussed limitations, in this paper, we propose to improve automated traceability recovery by eliciting transitive links from source, intermediate, and target artifacts with the help of consensual biterm extraction on artifact texts.
Specifically, we deduce the following two kinds of transitive links from different artifacts: (1) \textit{outer-transitive links}, and (2) \textit{inner-transitive links}. 
We deduce the former links from source to intermediate artifacts, and from intermediate to target artifacts while deducing the latter links within source artifacts and intermediate artifacts only.
We propose to use these two kinds of transitive links to concatenate relevant system functionalities across multiple artifacts and also to cluster the same system functionalities within each kind of non-target artifacts, respectively. 
We do not deduce inner-transitive links from target artifacts because their system functionalities are too fine-grained and diluted with too many implementation details.
We argue that \textit{our deduced inner- and outer-transitive links actually form implicit "networks" to comprehensively aggregate system functionalities across different software artifacts}, thus better bridging the semantic gap for automated tracing.

To deduce the inner- and outer-transitive links, we solely rely on textual similarities calculated by IR techniques because the artifact texts, though of which the quality and quantity are not satisfying, are still the prevailing perspective to analyze different software artifacts with heterogeneous structures.
This characteristic, along with the wide availability of artifact texts, is very beneficial for deducing transitive links across different kinds of source, intermediate, and target artifacts.
However, the performance of IR techniques on artifact texts is still vital for our approach.
Therefore, we introduce consensual biterm extraction \cite{Hui2022ASE} into our approach to enrich artifact texts and thus enhance the calculated IR values.
Specifically, we first extract an initial set of biterms for source, intermediate, and target artifacts separately.
Then, for biterms in source and target artifacts, we only retain those biterms that also appear in the intermediate artifacts.
We name the retained consensual biterms as \textit{intermediate-centric consensual biterms} because we argue that intermediate artifacts are naturally middle-level abstracted, and thus important to bridge the abstraction gap between source and target artifacts.
Based on the enriched texts of source and target artifacts with the retained biterms, we proposed to deduce inner-transitive links within each kind of artifact and outer-transitive links from source (or target) artifacts to intermediate artifacts when a given pair of artifacts has top-ranked textual similarities.
Finally, to improve automated traceability recovery based on IR techniques, we traverse the two deduced types of transitive links to form valid link paths and use these paths to adjust IR scores and improve the ranking of candidate trace links.
We will use the following example (adapted from the Dronology system \cite{dronologyDataset}) for further demonstration.

\begin{figure}[t]
    \centering
    \includegraphics[width=\linewidth]{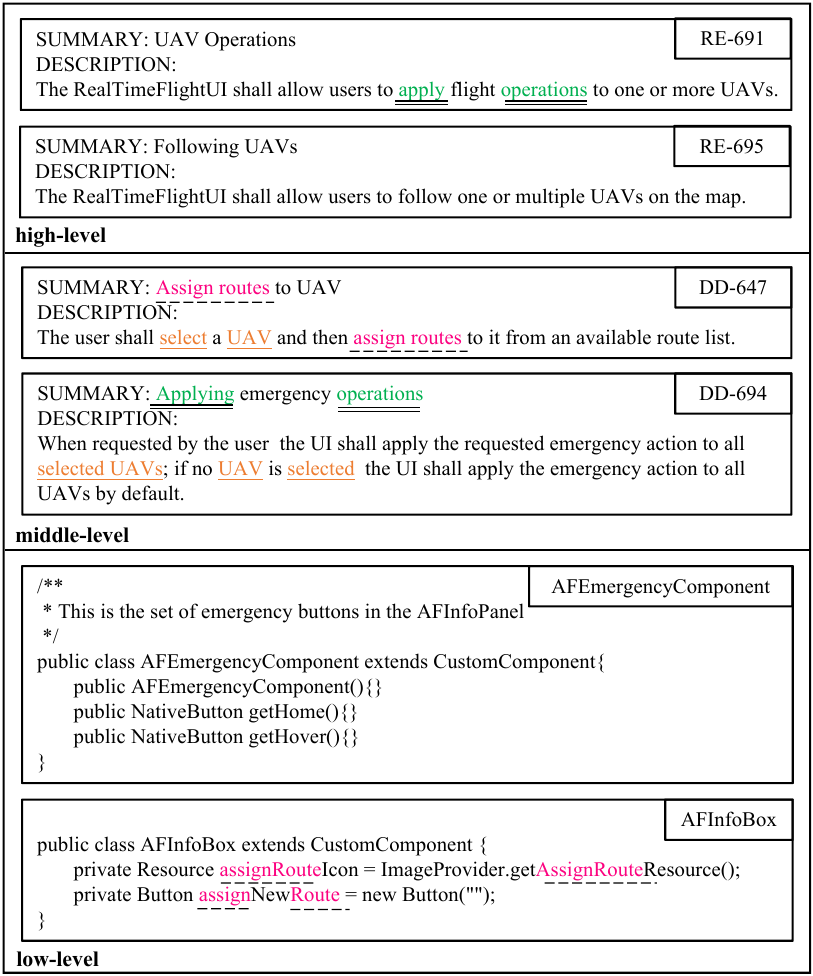}
    \caption{Motivating example adapted from the Dronology system with the following three consensual biterms: (\uuline{apply}, \uuline{operation}), (\dashuline{assign}, \dashuline{route}), and (\uline{select}, \uline{UAV})}
    \label{fig:case}
\end{figure}

Figure \ref{fig:case} shows five artifacts in different abstraction levels from an open-source Unmanned Aerial System called Dronology~\cite{DBLP:conf/icse/Cleland-HuangVB18}, including two high-level requirements (RE-691 and RE-695), two design definitions (DD-647 and DD-694), and the code snippets from class \texttt{AFInfoBox} and \texttt{AFEmergencyComponent}.
RE-691 describes the system functionality of ``UAV operations'', i.e., allowing users to apply flight operations to selected UAVs (Unmanned Aerial Vehicle). 
Accordingly, the class \texttt{AFInfoBox} is implemented to show UAV's information for flight operations, thus traced to RE-691.
Unfortunately, a significant abstraction gap can be observed through the texts of RE-691 and \texttt{AFInfoBox} where the two artifacts share few terms because RE-691 tends to describe the summarized concepts of ``UAV operations'', while \texttt{AFInfoBox} focuses more on the implementation details, such as showing the assigned routes for UAVs.
This situation hinders the automated recovery of the trace between RE-691 and \texttt{AFInfoBox} based on textual similarities, and we propose to deduce transitive links with the help of available intermediate artifacts and consensual biterm extraction.
Through the artifact texts with the additionally extracted consensual biterm (\uuline{apply}, \uuline{operation}), we can deduce an outer-transitive link between RE-691 and DD-694. Furthermore, with the help of extracted consensual biterm (\uline{select}, \uline{UAV}), we deduce an inner-transitive link between DD-694 and DD-647. 
Meanwhile, we can deduce another outer-transitive path between DD-647 and \texttt{AFInfoBox} mainly due to the extracted consensual biterm (\dashuline{assign}, \dashuline{route}).
Finally, we find a path from RE-691 to \texttt{AFInfoBox} based on the deduced two outer-transitive and one inner-transitive links.
We argue that this path can effectively bridge the previously observed abstraction gap between RE-691 and \texttt{AFInfoBox}, thus important for automated tracing the two artifacts.
Our proposed approach is discussed in Section \ref{sec:approach}.

% END motivating example
%-------------------------

We use five real-world systems to evaluate our approach, involving the three mainstream IR models (i.e., VSM, LSI, and JS). 
The results showed that our approach outperforms four state-of-the-art traceability recovery approaches, namely a naive IR approach without any enhancing strategies involved IR-ONLY, two IR-based approaches TAROT~\cite{Hui2022ASE} (using consensual biterms between requirements and code) and LIA~\cite{conf/icsm/RodriguezCF21} (leveraging intermediate artifacts), and a probabilistic-model-based approach COMET~\cite{DBLP:conf/icse/MoranPBMPSJ20}.
We implemented our approach called \textbf{TRIAD} (\textbf{T}raceability \textbf{R}ecovery 
by b\textbf{I}term-enh\textbf{A}nced \textbf{D}eduction of transitive links).
In summary, the main contributions of this work are:
(i) a novel approach called TRIAD using intermediate-centric consensual biterms and outer-inner combined transitive links for automated traceability recovery; 
(ii) an empirical evaluation of TRIAD on five open-source systems; 
and (iii) availability of source code and data of TRIAD online~\cite{TRIADcode, TRIADdataset}.

\section{Background and Related Work}
\label{sec:related}
This section discusses the background and related work on traceability recovery approaches, enhancing strategies for IR-based approaches, using biterms in lexical analyses, and leveraging multiple types of artifacts in traceability recovery.

\textbf{IR-based traceability recovery:} IR techniques are widely used for automated traceability recovery~\cite{DBLP:conf/icse/Cleland-HuangGHMZ14} because they can compute textual similarities between various software artifacts (e.g., requirements and code) based on the occurrence of terms from artifact texts, and suggest IR values as the probability whether a pair of source-target artifacts is a relevant trace.
Typically, IR-based approaches follow three steps to compute textual similarities ~\cite{DBLP:books/daglib/p/LuciaMOP12}: (i) constructing a corpus with terms extracted from different software artifacts; (ii) representing the corpus using a term-by-document matrix, where each artifact is organized as a document and cell values are weighted using the tf-idf weighting scheme ~\cite{Baezayates2004Modern} to emphasize the importance of each term in the document based on its frequency of occurrence; (iii) calculating the similarity among artifacts (represented as entries in the term-by-document matrix) using different IR models.
Current mainstream IR models include two vector-space-based VSM (Vector Space Model ~\cite{DBLP:journals/tse/AntoniolCCLM02}) and LSI (Latent Semantic Indexing ~\cite{DBLP:conf/icse/MarcusM03}), and one probability-based JS (The Jensen-Shannon model ~\cite{DBLP:conf/iwpc/AbadiNS08}). LSI is a variant of VSM and applies Singular Value Decomposition (SVD) ~\cite{Baezayates2004Modern} to the term-by-document matrix. 
To comprehensively evaluate TRIAD compared to IR-based approaches, our evaluation includes all three discussed IR models.

\textbf{Enhancement for IR-based approaches:} To address the vocabulary mismatch problem that greatly hinders the performance of IR-based traceability recovery, researchers have proposed various enhancing strategies from different perspectives, such as introducing enhanced lexical analyses on the text of artifacts~\cite{DBLP:conf/re/Cleland-HuangSDZ05, DBLP:conf/icsm/GethersOPL11,DBLP:conf/iwpc/LuciaPOPP11}, incorporating with execution tracing~\cite{DBLP:journals/tse/PoshyvanykGMAR07, DBLP:journals/ese/DitRP13}, or combining with the analyses on different kinds of code dependencies~\cite{DBLP:conf/icse/McMillanPR09,DBLP:conf/wcre/KuangNHRLEM17}. 
Furthermore, due to the semi-automatic nature of IR-based traceability recovery which requires user verifications on the generated candidate links, a different body of work~\cite{DBLP:journals/tse/HayesDS06,DBLP:conf/icsm/LuciaOS06, DBLP:conf/icse/PanichellaLZ15} uses user feedback (candidate links verified as either relevant links or false positives) to adjust the calculation of IR values. 
Panichella et al.~\cite{DBLP:conf/csmr/PanichellaMMPOPL13} further combined the user feedback with code dependency analysis, while Gao et al.~\cite{DBLP:journals/ese/GaoKMHLME22} proposed CLUSTER' that uses the closeness analysis on code dependencies~\cite{DBLP:conf/wcre/KuangNHRLEM17} to improve IR-based approaches by propagating only a small amount of user feedback. 
Unfortunately, although the discussed approaches achieved great progress, their improvements are still highly relevant to the quantity and quality of artifact texts, i.e., the initially calculated IR values.

\textbf{ML-based traceability recovery:} An increasing number of works proposed take advantage of Machine Learning (ML) techniques for traceability recovery and reported promising results, especially when some trace links have been previously identified, such as completing the missing links between issues and code commits in software repository~\cite{DBLP:conf/iwpc/LeVLP15, DBLP:conf/se/0002RGCM19, DBLP:journals/infsof/SunWY17}, or maintaining recovered traces through machine learning classification~\cite{conf/icsm/MillsEH18}. Mills et al.~\cite{conf/icsm/MillsEBKCH19} further introduced active learning to significantly reduce the amount of required training data for classification when recovering traces. 
However, these approaches still largely rely on calculated IR values as the features of their classification tasks, while our approach is expected to improve the quality of IR values, thus being likely to improve these approaches as well. 
Additionally, deep-learning-based approaches have been proposed, such as Guo et al.~\cite{DBLP:conf/icse/0004CC17} that uses the RNN network to generate links between subsystem requirements and design definitions, and T-BERT~\cite{DBLP:conf/icse/LinLZ0C21} that uses various BERT architectures to recover missing links between issues and commits. The deep learning (DL) network can capture the implicit connections from term sequences in artifact texts, whereas TRIAD explicitly deduces transitive links across source, intermediate, and target artifacts to bridge the abstraction gap without the demand of previously labeled traces as the training sets.

\textbf{Leveraging multiple types of software artifacts:}
Moran et al.\cite{DBLP:conf/icse/MoranPBMPSJ20} modeled the existence of traceability links based on probability, which incorporates user feedback, transitive links across groups of diverse development artifacts, and execution traces by running test cases (for only one evaluated system). 
However, they only extracted transitive relationships within requirements.  
Rodriguez et al.\cite{conf/icsm/RodriguezCF21} explored 84 kinds of different techniques combinations for leveraging intermediate artifacts to improve the accuracy of source-to-target trace recovery. 
However, they observed that it is hard to find a single technique that performed best across all datasets, while TRIAD achieves good performance by default settings, and also allows users to further fine-tune it (see Section \ref{sec:threats}).

\textbf{Using biterms in lexical analyses:} Biterms were originally proposed for document retrieval ~\cite{10.1145/564376.564476BTIR} to address the issue of data sparsity. 
For the same reason, Cheng et al.~\cite{6778764BTM} use biterms to better establish topic models over short texts. 
In the field of software engineering (SE) research, Hadi et al.~\cite{9240699} proposed an adaptive online biterm topic model to analyze version-sensitive short texts. 
Instead of using the biterm topic modeling directly, Gao et al.~\cite{Hui2022ASE} extracted consensual biterms from requirements and code to improve requirement traceability recovery.
However, they only considered the two specific types of artifacts, i.e., requirements and code. 
In contrast, our approach leverages multiple types of artifacts to extract intermediate-centric biterms for better deducing inner- and outer-transitive links, thus leading to a more versatile traceability recovery process by comprehensively bridging the abstraction gap.

% -- -- -- -- -- -- -- -- -- -- -- -- -- -- -- --
\begin{figure*}[!htp]
    \centering
    \includegraphics[width=\linewidth]{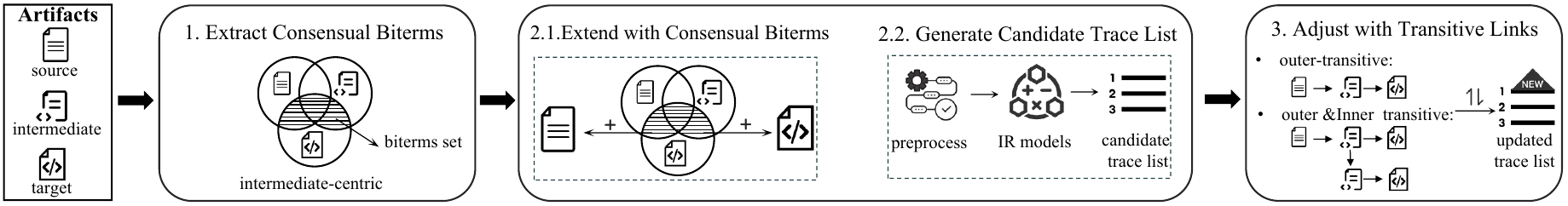}
    \caption{Overview of the TRIAD framework}
    \label{fig:triad}
\end{figure*}
\section{The TRIAD Approach}
\label{sec:approach} 
Figure~\ref{fig:triad} illustrates the overview of TRIAD.
First, we extract consensual biterms from source, intermediate, and target artifacts. 
Second, we enrich source and target artifact texts with consensual biterms from intermediate artifacts and generate candidate trace lists.
Finally, we further use inner- and outer-transitive links to adjust the candidate traces' IR scores. 
All steps are described in detail next.

\subsection{Extracting Intermediate-centric Biterms} \label{sec3.1}
Artifacts in our dataset are generally classified into two categories: natural language written artifacts (e.g., requirements, use cases, design definitions, etc.) and programming language artifacts (e.g., source code and test code). 
We then introduce two separate biterm-extraction strategies for the two categories of artifacts according to Gao et al.'s work ~\cite{Hui2022ASE}.

\begin{figure}[]
    \centering
    \includegraphics[width=\linewidth]{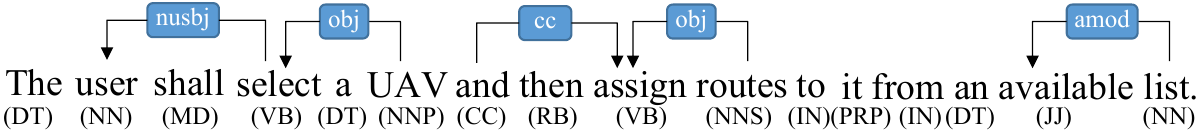}
    \caption{Stanford CoreNLP parse result for the sentence in DD-647.}
    \label{fig:sentence}
\end{figure}

\subsubsection{Extracting Biterms from Natural Language Artifacts:}\label{sec3.1.1}
For natural language artifacts, we leverage Stanford CoreNLP~\cite{DBLP:conf/acl/ManningSBFBM14} to parse the texts with steps as follows: 
(i) split a text into independent sentences; 
(ii) use Part-Of-Speech (POS, e.g., nouns, verbs, adjectives, adverbs, pronouns, etc.) Tagger\footnote{https://nlp.stanford.edu/software/tagger.html} to assign POS to each term in a sentence; 
(iii) use Stanford-typed dependencies parser\footnote{https://nlp.stanford.edu/software/stanford-dependencies.html} on each sentence to get term pairs (biterms) having grammatical relationship in the sentence; 
and (iv) filter biterms that are not composed of nouns, verbs, or adjectives.
For example, Figure~\ref{fig:sentence} shows the parse result of the given sentence in the design definition DD-647.
The POS tag of each term is provided in the brackets, including NNP (proper noun), MD (modal auxiliary), VB (verb), NN (noun), NNS (noun, plural), IN (preposition), CC (coordinating conjunction), and JJ (adjective).
In this example, we obtain five biterms with grammatical relationships, i.e., \textbf{select, user} (\textit{nusbj}, nominal subject), \textbf{select, UAV} (\textit{obj}, object), \textbf{assign, and} (cc, coordination), \textbf{assign, routes} (\textit{obj}, object), and \textbf{availiable, list} (\textit{amod}, adjective modifier).
Then we discard \textbf{assign, and} since it does not satisfy the POS combination criterion. 
The remaining four biterms are considered as candidate biterms after preprocessing and saved as \textbf{(user, select)}, \textbf{(select, uav)}, \textbf{(assign, rout)}, and \textbf{(avail, list)}.

\subsubsection{Extracting Biterms From Programming Language Artifacts:} \label{sec3.1.2}
For programming language artifacts, we extract biterms from code identifier names (e.g., class name, method name, invoked method name, field type and name, parameter type and name, etc.) and comments (i.e., comments for class or method). 
To identify these code elements, we use srcML~\cite{7203124, srcML} to translate Java and C source code to the srcML format, which wraps the text of the source code in XML elements. 
Hence, it allows us to identify and extract these code elements precisely. 
Then, the extraction steps are: (i) extract code identifiers and comment data from the source code; (ii) extract consensual biterms from preprocessed code identifiers and comments.
Currently, existing POS taggers cannot achieve 100\% accuracy in tagging code identifiers due to a lack of sentence context and grammar structures \cite{Newman-jss2020}.
We apply the same extraction strategies for identifier names by combining any two split terms sequentially proposed in TAROT \cite{Hui2022ASE}.
For example, for class \texttt{AFInfoBox} in the motivation case, we can get \textbf{(af, info)}, \textbf{(af, box)}, and \textbf{(info, box)} from its class name.
As for its field name \texttt{assignRouteIcon}, we can get \textbf{(assign, rout)}, \textbf{(assign, icon)}, and \textbf{(rout, icon)}.
Furthermore, we can also get additional two occurrences of \textbf{(assign, rout)} from its invoked method name \texttt{getAssignRouteResource} and another field name \texttt{assignNewRoute} respectively.

\subsubsection{Retaining Intermediate-centric Consensual Biterms:} \label{sec3.1.3}
Once we have extracted biterms from the source, intermediate, and target artifacts, we filter them to obtain intermediate-centric consensual biterms. 
Specifically, biterms extracted from the source and target artifacts must also appear in any of the intermediate artifacts.
Furthermore, we discard biterms that appear only in the intermediate artifacts. 
We argue that these biterms would introduce noise during textual similarity calculation since they do not appear in either the source or the target artifacts.
For example, for class \texttt{AFInfoBox} in the motivation example, we extract (af, info), (af, box), (info, box), (assign, route), (assign, icon), and (rout, icon).
After the filter process, only (assign, rout) will be retained because other biterms do not appear in intermediate artifacts (i.e., DD-647 and DD-694).

\subsection{Enriching Artifacts with Biterms and Calculating IR Candidate List} \label{sec3.2}
\subsubsection{Enriching Source, Intermediate, and Target artifacts with \linebreak Biterms:} \label{sec3.2.1}
After extracting biterms from source, intermediate, and target artifacts, we subsequently determine their occurrence frequency as an indicator of importance.
For natural language artifacts, the importance of a biterm is determined directly by how often that biterm occurs in the entire text of the underlying artifact.
For programming language artifacts, we follow the rules proposed in TAROT \cite{Hui2022ASE}, which consider different parts of code of different importance and are defined as follows.
For a biterm that appears in class names or method names, its importance count is increased by two, since the names of classes and methods are particularly important~\cite{DBLP:conf/ecoop/HostO09}. 
A biterm that appears in comments (i.e., comments of classes and methods) has its count increased by one for each occurrence.
For invoked method names, field types and names, and parameter types and names, if the biterm only appears in these parts, its importance count is assigned to one and remains unchanged regardless of how many times it appears.
Finally, for each biterm we accumulate all importance counts from the various code parts. 
For example, a biterm that appears once in the class name, twice in the comments, and three times in the parameter types obtains an importance count of $1\times2+2\times1+1=5$. 
\subsubsection{Creating Corpus and Computing Textual Similarities:}\label{sec3.2.2}
Each programming written artifact is extracted into one document containing its comments and identifiers, including class names, method names, field types and names, and parameter types and names.
Then we use the CamelCase and snake\_case naming conventions to split identifiers into terms. 
For natural language written artifacts, we extract a document that includes all content. 
The documents are normalized by standard pre-processing techniques for IR, by: removing special tokens, converting upper case letters into lower case, removing stop words, and performing the Porter stemming algorithm~\cite{DBLP:journals/program/Porter80}.
Finally, we calculate textual similarities based on three mainstream IR models: Vector Space Model (VSM)~\cite{DBLP:journals/tse/AntoniolCCLM02}, Latent Semantic Indexing (LSI)~\cite{DBLP:conf/icse/MarcusM03}, and the probabilistic Jensen and Shannon (JS) model~\cite{DBLP:conf/iwpc/AbadiNS08}.
\subsubsection{Enriching Source and Target artifacts with Intermediate-centric Biterms} \label{sec3.2.3}
First, we identify highly related intermediate artifacts for each source and target artifact based on their textual similarities. 
Using the similarity scores calculated in Step 2 we only consider at most $t$ intermediate artifacts, where $t$ = 3 and for each artifact the similarity score must be equal to or greater than threshold $m$ = 0.5 of the maximum similarity (thresholds calibration discussed in Section \ref{sec3.3.3}).
In other words, a requirement with the top most similar design definition having a similarity score of 0.8, results in considering two further design definitions as long as their similarity score is higher than 0.4 (= $0.8\times 0.5$).
Then we use the consensual biterms extracted from the highly related intermediate artifacts to enrich the source and target artifacts, whereby each consensual biterm is added only once. 
For example, requirement RE-691 can be extended with (appl, oper)and (select, uav) from DD-694.
Class \texttt{AFInfoBox} can be extended with (select, uav) and (assign, rout) from DD-647.

\subsubsection{Generating Candidate Links:}\label{sec3.2.4}
We pairwise calculate the textual similarities between enriched source, target, and intermediate artifacts.
For each source artifact, we obtain a list of ranked IR target candidates. 
The list is sorted in descending order determined by the textual similarity using three mainstream IR models (see Sec.\ref{sec3.2.2}).

\subsection{Adjusting IR Scores by Transitive Links} \label{sec3.3}
% \vspace{2mm} \noindent

Intermediate artifacts provide not only intermediate-centric consensual biterms, but also transitive semantics from source to target artifacts.
We deduce two types of transitive links based on textual similarities: between the same kind (\textit{inner-transitive}) and between different kinds of artifacts (\textit{outer-transitive}), as described next.

\subsubsection{Outer-transitive Links:} \label{sec3.3.1}
\begin{figure}[!bp]
    \centering
    \includegraphics[width=.8\linewidth]{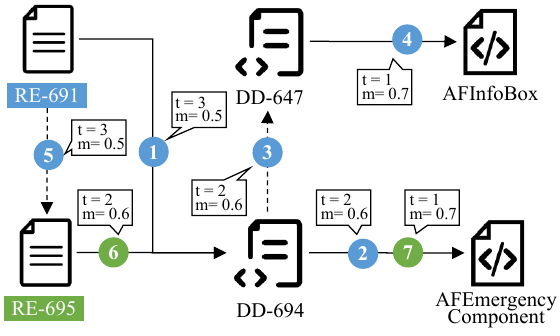}
    \caption{Applying TRIAD in the motivating scenario depicted in Figure \ref{fig:case}}  \label{fig:approach-case}
\end{figure}

We deduce an outer-transitive link between a pair of different kinds of artifacts (at different abstraction levels) that exhibit high textual similarities, thus likely to be semantically correlated. 
That is to say, if a source artifact \texttt{S} has a high similarity with an intermediate artifact \texttt{I}, and \texttt{I} also has a high similarity with a target artifact \texttt{T}, then we assume that \texttt{S} is closely related to \texttt{T}, and the outer-transitive link is \texttt{S}→\texttt{I}→\texttt{T}.
Note that we do not rely on any explicit trace links that might exist between source to intermediary artifact or target to intermediary artifact as such traces might not always be available.
For example, as shown in Figure \ref{fig:approach-case} there exists an outer-transitive link, i.e., RE-691→DD-694→\texttt{AFEmergencyComponent}.
The outer transitive links are connected with a solid black arrow.
Through this transitive link, we can bridge the abstraction gap between high-level abstracted artifact RE-691 and low-level abstracted code file \texttt{AFEmergencyComponent}. 

\subsubsection{Inner-transitive Links:} \label{sec3.3.2}
We deduce an inner-transitive link between any two artifacts with high textual similarity that are of the same type as they reside at the same level of abstraction. 
We assume that if a source artifact \texttt{S$_{1}$} has a high similarity with another source artifact \texttt{S$_{2}$}, there is an inner-transitive link \texttt{S$_{1}$}→\texttt{S$_{2}$}.
For example, as shown in Figure \ref{fig:approach-case} there are two inner-transitive links connected with dashed black arrows, i.e., RE-691→RE-695 and DD-694→DD-647.
As we discussed in Section \ref{sec:introduction}, we do not deduce inner-transitive links from target artifacts due to their too fine-grained system functionalities diluted by implementation details. Certain target artifacts such as code are highly structured and contain inherent dependencies such as calling relationships. However, we do not consider these inner dependencies in our approach either, even when they are available and have high quality. The reason is that introducing target-level inner-transitive links will lead to too many possible paths when traversing all transitive links, even though these links are rigorously chosen based on textual similarities. This situation will over-aggregate the system functionalities and eventually decrease the performance of automated tracing, which is also confirmed during our trials and experiments.

\subsubsection{Filtering Inner- and Outer-transitive Links:}\label{sec3.3.3}
We now discuss how to choose outer- and inner-transitive links based on textual similarities.
We apply the same two thresholds introduced above: (i) $m$ (=0.5) ensures that the selected transitive links' IR values are equal to or greater than 50\% of the maximum similarity, which is determined by the most related intermediate artifact.
By considering this threshold, we aim to prioritize transitive links with higher similarity scores and disregard those with lower ranks that potentially indicate lower similarity or non-relevance.
(ii) $t$ (=3) represents the maximum number of outer- and inner-transitive links to be selected, respectively.
We adopt a conservative approach by considering only the top $t$ transitive links to avoid introducing too many intermediate artifacts, due to enriching artifacts with overmuch biterms that may dilute their unique semantics.
Furthermore, we consider the decreasing reliability of textual similarity as we include more artifacts in the transitive links. 
To mitigate this risk, we adopt a specific strategy. 
For $m$, we increase the threshold by $0.1$ for every hop from the source artifact ($m'= 0.1\times n + m$, where
$n$ is the number of hops). 
Conversely, for the threshold $t$ we reduce $t$ by $1$ for each hop. %by the current path number $n$. 
This approach aims to avoid introducing potentially noisy transitive links.
For practical recommendations of threshold calibration, please refer to our further discussions in Section~\ref{sec:threats} (the "Implications for Practitioners" paragraph).

\subsubsection{Forming Paths from Transitive Links to Adjust IR Scores:}\label{sec3.3.4}
Now we can further combine inner-transitive links with outer-transitive links to find the potential relationship between artifacts.
That is to say, in Figure \ref{fig:approach-case}, if \texttt{S$_{2}$} has a high similarity with an intermediate artifact \texttt{I}, there is likely a relation between \texttt{S$_{1}$} and \texttt{I}.
We can find the inner-outer-transitive link is \texttt{S$_{1}$}→\texttt{S$_{2}$}→\texttt{I}.
For example, RE-691→RE-695→DD-694→\texttt{AFEmergencyComponent} is an inner-outer-transitive link in Figure \ref{fig:approach-case}.
Again, we do not rely on explicit, existing trace links between artifacts of the same type as such traces might not be available in every environment and for every artifact type.
Figure \ref{fig:approach-case} further illustrates the adjustment process for the motivating scenario. 
Starting with the source artifact RE-691, we calculate the candidate list of intermediate artifacts (i.e., outer-transitive links). 
We find that DD-694 has the highest similarity \textcircled{1}, and the other intermediate artifacts do not satisfy the $m$ (=0.5). 
Therefore, we select DD-694 as the intermediate artifact and compute similarities between DD-694 and the target artifacts.
Considering the existing path RE-691→DD-694 labeled with \textcircled{1} (i.e., constituting one hop), we reduce $t$ by 1 and add 0.1 to $m$. 
Among the target artifacts, \texttt{AFEmergencyComponent} has the highest similarity, while the others fall below the adjusted $m$ (0.5 + 0.1). Thus, we identify an outer-transitive link RE-691→DD-694→\texttt{AFEmergencyComponent}, and label the related paths as \textcircled{1} and \textcircled{2}.
Next, we calculate the similarities between DD-694 and other intermediate artifacts to find the inner-transitive links. 
Given the current $t$ (=2) and adjusted $m$ (=0.6), which resulted from including the path RE-691→DD-694 in the transitive link, we discover that DD-647 satisfies both threshold constraints. 
As a result, we can further use the outer-transitive relation of DD-647 to identify the target artifacts. 
At this stage, the $t$ is 1, and $m$ is 0.7 due to two hops included in the transitive link (RE-691→DD-694 \textcircled{1} and DD-694→DD-647 \textcircled{3}).
Finally, only \texttt{AFInfoBox} exceeds thresholds and we can find an outer-inner-transitive link RE-691→DD-694→DD-647→\texttt{AFInfoBox} labeled with numbers \textcircled{1}\textcircled{3}\textcircled{4}.

It is worthwhile noting that for each inner-outer-transitive link, our approach only allows the link to contain one inner-transitive link.
The reason is that we found that two different artifacts from the same abstraction level tend to represent different parts of the system functionalities, i.e., the two sets of their traced, different abstraction-level artifacts are more likely to have a small overlap only.
Hence, introducing more than one inner-transitive links may lead to the inclusion of more potentially unrelated artifacts in the deduction phase, especially when the influence of each inner-transitive link is further amplified by the correlated outer-transitive links from each inner-outer-transitive link, thus decreasing the precision of TRIAD.
For instance, because there is already an inner-transitive link RE-691→RE-695 \textcircled{5}, the transitive link RE-691→RE-695→DD-694 \textcircled{5}\textcircled{6} will not include another inner-transitive link DD-694→DD-647 \textcircled{3} but only a direct hop to the destination artifact \textcircled{7}.
Consequently, all transitive links are of hop count two when comprising only outer-transitive links or of hop count three when containing one inner-transitive link.
In Figure \ref{fig:approach-case}, all the transitive links are sequentially labeled with numbers based on their execution order.
Links originating from RE-691 and RE-695 are distinguished with blue and green colors, respectively.

After identifying the transitive links, we proceed to calculate the $bonus$ by multiplying the IR score of each link: $bonus = \prod_{i=1}^{n}{IR_{link_i}}$. 
Next, we utilize the $bonus$ value to enhance the initial IR score $IR$ between the source and target artifact along this transitive path. 
The updated IR score $IR^{'}$ is determined as follows:
$IR^{'} = IR \times (1 + bonus)$.

% -- -- -- -- -- -- -- -- -- -- -- -- -- -- -- --
\section{Experiment Setup}
\label{sec:setup}
To evaluate TRAID, we perform an empirical evaluation with two major goals:
(i) assess whether biterm-enhanced deduction of transitive links can improve IR-based traceability recovery, and (ii) investigate the individual impacts of biterms, outer-, and inner-transitive links on TRIAD.
We formulate the following two RQs:

\vspace{2mm} \noindent
\textbf{RQ1: \textit{To what extent does TRIAD exceed the performance of baseline approaches?}}

\vspace{2mm} \noindent
\textbf{RQ2: \textit{What is the individual impact of biterms, outer- and inner-transitive on performance?}}

\vspace{2mm} \noindent
To answer our two research questions, we conducted several experiments with four baseline approaches and five evaluated systems.

\subsection{Evaluation Procedures}
\subsubsection{RQ1: Comparing TRIAD Against Four Baselines}
To answer RQ1, we compare the performance of TRIAD with four baseline approaches.
We first choose \textbf{IR-ONLY}, the naive IR approach without any enhancing strategies involved.
Then considering that TRIAD is based on biterm-enhanced deduction of transitive links among multiple artifacts, we select a biterm-based approach \textbf{TAROT} ~\cite{Hui2022ASE} and two approaches using multiple types of artifacts, i.e, \textbf{LIA} ~\cite{conf/icsm/RodriguezCF21} and \textbf{COMET} ~\cite{DBLP:conf/icse/MoranPBMPSJ20} as follow: 
(1) TAROT enriches texts of artifacts with consensual biterms extracted from two types of structured artifacts, i.e., requirement and code. 
Note that we do not use the local weight $\theta$ proposed in TAROT, which considers that biterms extracted from different structure parts of artifacts should be given different weights. 
However, evaluated systems in our experiment have various types of artifacts, with different structures or even without apparent structures.
(2) LIA is a approach that \textbf{L}everages \textbf{I}ntermediate \textbf{A}rtifacts to improve IR-based traceability recovery. 
LIA involves 84 strategy combinations, however, its authors acknowledged that there is no single technique that performs the best across all datasets.
Then, we used two variants of this approach: (i) LIA$_{rsc}$ with \textbf{r}ecommended \textbf{s}trategy \textbf{c}ombination, with a global scaling for link score normalization, sum of link aggregation, and sum for aggregating direct and transitive results, which tries to reach good performance on all datasets; and (ii) LIA$_{best}$, which performs best across all the variants using the same IR model. Notice that LIA only supports VSM and LSI. 
(3) COMET models traceability problems from the probability perspective, incorporating user feedback and transitive relationships. 
Note that COMET uses a unified textual similarity computed by a set of IR/ML techniques, so we cannot compare TRIAD with COMET based on a single IR model. 
To remain fair and comprehensively compared with COMET, we choose TRIAD with the lowest and highest average precision (AP) performances across the three IR models, since the authors that proposed COMET only report AP in their work.
Furthermore, we compare our results to two versions of holistic COMET (including developer feedback and transitive links), COMET$_{map}$ using Maximum a Posteriori (MAP) estimation \cite{BassettD19} and COMET$_{nuts}$ using a Markov Chain Monte Carlo (MCMC) technique via the No-U-Turn sampling (NUTS) process \cite{HoffmanG14}.

To compare our approach with baseline approaches, we use the \mbox{F-measure} at each recall point as the single dependent variable of our significant test. Then, to check the statistical difference between the results, we applied the Wilcoxon Rank Sum test~\cite{WilcoxonIndividual}, following null hypothesis: \textit{$H_0$: There is no difference between the performance of TRIAD and baseline approaches.}
We use the customary $\alpha$ = 0.05 of significance level (i.e., 95\% of confidence) to accept or refute $H_0$. 

Additionally, to quantify the difference between TRIAD and the baseline approaches, we applied the Cliff's delta ($\delta$) effect size~\cite{MacbethCliff}, as follows:
\begin{equation}
	\delta=\left |  \frac{\#\left ( x_{1}> x_{2}\right ) - \#\left ( x_{1}<  x_{2}\right )}{n_{1}n_{2}}   \right |
\end{equation}

where $x_1$ and $x_2$ represent F-measures of TRIAD and a chosen baseline approach, and $n_1$ and $n_2$ are the sizes of the sample groups. 
The effect size is considered 
negligible for $\delta < 0.15$, 
small for $0.15 \leq \delta < 0.33$, 
medium for $0.33 \leq \delta < 0.47$, 
large for $\delta > 0.47$.

\subsubsection{RQ2: Ablation Study}
To answer RQ2, we conduct an ablation study to evaluate the performance of different parts of TRIAD.
We started with the IR-ONLY approach and incrementally added each part of TRIAD to evaluate its impact. 
By comparing the performance before and after adding each part, we measured the influence of individual parts of TRIAD.
Initially, we utilized only the intermediate-centric consensual biterms (denoted as ``\textbf{b}'') to enhance the input quality for IR techniques. 
Subsequently, we introduced the \textit{outer-transitive links} to adjust the IR scores through the biterms (denoted as ``\textbf{b+o}''), and later incorporated the \textit{inner-transitive links} as well (denoted as ``\textbf{b+o+i}'').
Moreover, we want to explore the performance of using transitive links independently, without combining them with biterms. 
Therefore, we conducted TRIAD runs exclusively with the outer-transitive links (denoted as ``\textbf{o}'') and with both outer- and inner-transitive links (denoted as ``\textbf{o+i}'').

\begin{table*}[tb]
\caption{Overview of the Five Evaluated Systems (S = Source, I = Intermediate, T = Target)}
\label{tab:dataset}
\addtolength{\tabcolsep}{-2pt}
\centering
\resizebox{.9\textwidth}{!}{
\tiny
% \resizebox{\textwidth}{!}{
\begin{tabular}{l |l|l|l |l|l|l}
\hline%\toprule
\multirow{2}{*}{\textbf{Dataset}} & \multicolumn{3}{c|}{\textbf{Artifacts}} & \multicolumn{3}{c}{\textbf{Traces}}\\
\cline{2-7}%\cmidrule{2-7}
 & \textbf{Source} & \textbf{Intermediate} & \textbf{Target}& \textbf{S→I} & \textbf{I→T} & \textbf{S→T}\\
\hline%\midrule 
Dronology & Requirement:58 & Design Definitions:144 & Source code:184 & Req→DD:132 & DD→Src:563 & Req→Src:393\\

WARC & Non-Func. Reqs:21 & Specifications:89 & Func. Reqs:42 & NFR→SRS:58 & SRS→FRS:78 & NFR→FRS:45\\

EasyClinic& Use Case:30  & Interaction Description.:20 & Code Descr.:47 & UC→ID:132 & ID→CD:563 & UC→CD:393\\

EBT& Requirement:44 & Test Case Description.:25 & Source code:50 &Req→TC:51 & TC→Src:93 & Req→Src:98\\

LibEST & Requirement:52  & Test Code:21 & Source code:14 & Req→Test:352 & Test→Code:108 & Req→Code:204\\
\hline%\bottomrule
\end{tabular}}
\end{table*}

\subsection{Evaluation Metrics}  \label{sec5.2}
To evaluate the performance of our approach, we rely on five well-known metrics, namely \textit{Precision}, \textit{Recall}, \textit{F-measure}, \textit{Average Precision}, and \textit{Mean Average Precision}.

\textit{Precision} represents the proportion of correct links among retrieved trace links, calculated as follows:
\begin{equation}
	Precision = \frac{\left | relevant\cap retrieved\right |}{\left | retrieved\right |}\%
    \end{equation}

\textit{Recall} represents the proportion of retrieved correct links among all correct links, calculated as follows:
    \begin{equation}
	Recall = \frac{\left | relevant\cap retrieved\right |}{\left | relevant\right |}\%      
    \end{equation}
where $relevant$ is the set of relevant links and $retrieved$ is the set of links retrieved by traceability recovery approaches.

\textit{F-measure (F)} represents the harmonic mean of precision and recall values, calculated as follows:
\begin{equation}
	F=\frac{2Precision \times Recall}{Precision + Recall}
\end{equation}
where \textit{P} represents precision, \textit{R} represents recall, and \textit{F} is the harmonic mean of \textit{P} and \textit{R}. 
A higher F-measure means that both precision and recall are high and balanced. 

\textit{Average Precision (AP)} measures how well relevant documents of all queries (requirements) are ranked to the top of the retrieved links, calculated
as follows:
    \begin{equation}
	AP = \frac{\sum_{r=1}^{N}(Precision(r)\times isRelevant(r))}{\left | RelevantDocuments\right |}
    \end{equation}
where $r$ is the rank of the target artifact in an ordered list of links, $Precision(r)$ represents its precision value, $isRelevant()$ is a binary function assigned 1 if the link is relevant or 0 otherwise, $N$ is the number of all documents. 
    
\textit{Mean Average Precision (MAP)} is the average of the AP scores of all queries to measure how well relevant documents for each query are ranked to the top of the retrieved links, calculated as follows:
\begin{equation}
	MAP = \frac{\sum_{q=1}^{Q}AP(q)}{Q}
\end{equation}
where $q$ is a single query, and $Q$ is the number of all queries.

\subsection{Evaluated Systems}  \label{sec5.1}
Our evaluation is based on five real-world software systems: Dronology, WARC, EasyClinic, EBT, and Libest. 
Table~\ref{tab:dataset} presents a summary of five evaluated systems.
Dronology is an open-source Unmanned Aerial System.
We collect its dataset from their website~\cite{dronologyDataset}, retaining only requirements and design definitions in closed status and traced code files. 
EasyClinic, WARC, and EBT are from the open source datasets from the CoEST community~\cite{CoESTDataset}. 
LibEST from Cisco is an open-source EST stack written in C and used for secure certificate enrollment~\cite{LibESTDataset}. 
It should be noted that to compare with the baseline approach COMET~\cite{DBLP:conf/icse/MoranPBMPSJ20} on EBT and LibEST without bias, we use their processed data and result lists published in their dataset.
In the work that presents the baseline LIA~\cite{conf/icsm/RodriguezCF21} (Leverage intermediate artifacts), the authors built their ground truth by using the Connecting Links Method (CML) proposed by Nishikawa et al.~\cite{conf/icsm/NishikawaWFOM15}.
However, we noted that the results shown in CLM indicate it can only gain a slight improvement and cannot apply to all data, thus we only use the ground truth defined in the dataset to avoid introducing unpredictable factors.

\begin{table*}[tb]
\caption{The number of computed AP, MAP, p-value, and Cliff’s $\delta$ evaluating each approach (RQ1)}
\renewcommand\arraystretch{0.95}
\setlength{\tabcolsep}{4pt}
\centering
% \resizebox{\textwidth}{!}{
\resizebox{.9\textwidth}{!}{
\tiny
\label{tab:RQ1}
\begin{tabular} {l|l|llll|llll|llll}
\hline%\toprule

&      & \multicolumn{4}{c}{VSM} & \multicolumn{4}{c}{LSI} & \multicolumn{4}{c}{JS} \\ 
\cline{3-6} 	\cline{7-10}	\cline{11-14}
&          &AP& MAP    & $p$-value 	  & C'$\delta$    
           &AP& MAP    & $p$-value    & C'$\delta$       
           &AP& MAP    & $p$-value    & C'$\delta$    \\ 

\hline%\midrule		
\multirow{6}*{Dronology}  
& IR-ONLY	      & 18.93 & 35.88 & \textbf{\textless{}0.01} & 0.11 
                  & \textbf{22.74} & 40.75 & \textbf{0.01}            & 0.10  
                  & 17.63 & 34.38 & \textbf{\textless{}0.01} & 0.14 \\
& TAROT           & 18.70 & 38.97 & \textbf{0.03}            & 0.09 
                  & 19.28 & 42.19 & \textbf{0.04}            & 0.08 
                  & 17.19 & 39.05 & \textbf{\textless{}0.01} & 0.12 \\
& LIA$_{rsc}$     & 14.64 & 35.49 & \textbf{\textless{}0.01} & 0.13 
                  & 15.76 & 35.59 & \textbf{\textless{}0.01} & 0.25 
                  & -     & -     & -               & -    \\
& TRIAD           & \textbf{20.07} & \textbf{43.25} & -               & -    
                  & 21.13 & \textbf{49.07} & -               & -    
                  & \textbf{20.55} & \textbf{46.39} & -               & -    \\
% \cmidrule(r){3-14}
\cline{2-14}
& LIA$_{best}$    & 16.83 & 41.46 & 0.25            & 0.05 
                  & 20.23 & 41.68 & 0.13            & 0.06 
                  & -     & -     & -               & -    \\
\hline%\midrule

\multirow{6}*{WARC}  
& IR-ONLY		  & 41.41 & 65.01 & 0.23            & 0.15 
                  & 44.20 & 61.24 & \textbf{0.03}            & 0.27 
                  & 43.61 & 66.63 & 0.20            & 0.16 \\
& TAROT           & 43.62 & 66.58 & 0.25            & 0.14 
                  & 35.73 & 51.77 & \textbf{\textless{}0.01} & 0.45 
                  & 44.30 & 66.54 & 0.15            & 0.18 \\
& LIA$_{rsc}$	  & 43.64 & \textbf{72.98} & 0.76            & 0.04 
                  & 34.54 & 59.96 & \textbf{\textless{}0.01} & 0.49 
                  & -     & -     & -               & -    \\
& TRIAD           & \textbf{45.74} & 70.49 & -               & -    
                  & \textbf{51.26} & \textbf{63.77} & -               & -    
                  & \textbf{46.21} & \textbf{71.21} & -               & -    \\
% \cmidrule(r){3-14}
\cline{2-14}
& LIA$_{best}$    & 45.38 & 74.55 & 0.44            & 0.10 
                  & 47.07 & 71.41 & 0.18            & 0.16 
                  & -     & -     & -               & -    \\
\hline%\midrule

\multirow{6}*{EasyClinic}  
& IR-ONLY		  & 65.39 & 76.51 & 0.22            & 0.10 
                  & 53.51 & 69.47 & \textbf{0.01}            & 0.21 
                  & 46.75 & 62.67 & \textbf{0.01}            & 0.21 \\
& TAROT           & 60.19 & 69.24 & \textbf{\textless{}0.01} & 0.24 
                  & 52.24 & 67.15 & \textbf{\textless{}0.01} & 0.30 
                  & 46.90 & 57.69 & \textbf{\textless{}0.01} & 0.36 \\
& LIA$_{rsc}$     & 55.75 & 74.02 & \textbf{0.01}            & 0.21 
                  & 52.29 & 71.25 & \textbf{\textless{}0.01} & 0.31 
                  & -     & -     & -               & -    \\
& TRIAD           & \textbf{68.32} & \textbf{79.05} & -               & -    
                  & \textbf{64.55} & \textbf{78.01} & -               & -    
                  & \textbf{56.35} & \textbf{69.10} & -               & -    \\
% \cmidrule(r){3-14}
\cline{2-14}
& LIA$_{best}$    & 71.91 & 82.05 & 0.93            & 0.01 
                  & 70.37 & 79.81 & 0.77            & 0.02 
                  & -     & -     & -               & -    \\
\hline%\midrule

\multirow{6}*{EBT}  
& IR-ONLY		  & \textbf{23.25} & 33.03 & \textbf{0.02}            & 0.20 
                  & 20.33 & 39.44 & \textbf{\textless{}0.01} & 0.31 
                  & 20.81 & 33.39 & \textbf{\textless{}0.01} & 0.26 \\
& TAROT           & 20.42 & 33.66 & 0.83            & 0.02 
                  & 17.00 & 37.52 & \textbf{\textless{}0.01} & 0.33 
                  & 16.74 & 34.73 & \textbf{0.03}            & 0.18 \\
& LIA$_{rsc}$	  & 17.32 & 35.87 & 0.41            & 0.07 
                  & 15.54 & 37.61 & \textbf{\textless{}0.01} & 0.31 
                  & -     & -     & -               & -    \\   
& TRIAD           & 23.04 & \textbf{38.10} & -               & -    
                  & \textbf{22.30} & \textbf{40.55} & -               & -    
                  & \textbf{21.00} & \textbf{39.68} & -               & -    \\
% \cmidrule(r){3-14}
\cline{2-14}
& LIA$_{best}$    & 14.86 & 37.27 & 0.25            & 0.10 
                  & 16.61 & 38.01 & \textbf{\textless{}0.01} & 0.38 
                  & -     & -     & -               & -    \\
\hline%\midrule

\multirow{6}*{LibEST}  
& IR-ONLY		  & 55.25 & 73.30 & 0.21            & 0.07 
                  & 50.94 & 62.54 & 0.07            & 0.10 
                  & 57.69 & 66.63 & \textbf{\textless{}0.01} & 0.17 \\
& TAROT           & 57.01 & 73.52 & 0.85            & 0.01 
                  & 47.67 & 64.83 & 0.50            & 0.04 
                  & 61.31 & 73.43 & \textbf{\textless{}0.01} & 0.15 \\
& LIA$_{rsc}$	  & 52.06 & \textbf{77.36} & 0.60            & 0.03 
                  & 51.70 & \textbf{76.16} & 0.69            & 0.02 
                  & -     & -     & -               & -    \\
& TRIAD           & \textbf{56.86} & 72.40 & -               & -    
                  & \textbf{55.62} & 69.81 & -               & -    
                  & 59.13 & \textbf{75.26} & -               & -    \\
% \cmidrule(r){3-14}
\cline{2-14}
& LIA$_{best}$    & 50.35 & 77.67 & \textbf{0.03}            & 0.13 
                  & 51.70 & 76.16 & 0.70            & 0.02 
                  & -     & -     & -               & -   \\
		
\hline%\bottomrule

\end{tabular}}
\end{table*}

\begin{table}[tb]
\label{tab:comet}
\caption{Comparing COMET and TRIAD with metrics AP and MAP on systems EBT and LibEST}
\centering
\tiny
\resizebox{.9\linewidth}{!}{
\begin{tabular}{l|l|l|l|l}
\hline
            & \multicolumn{2}{c}{EBT} & \multicolumn{2}{c}{LibEST} \\
            \cline{2-5}
            & AP         & MAP        & AP           & MAP         \\
            \hline	
COMET$_{map}$    & 15.82      & 33.70       & 63.13        & 74.31       \\
COMET$_{nuts}$   & 16.60       & 33.43      & \textbf{63.45}        & 74.10        \\
TRIAD$_{low}$ & 21.00         & \textbf{39.68}      & 55.62        & 69.81       \\
TRIAD$_{high}$   & \textbf{23.04}      & 38.10       & 59.13        & \textbf{75.26}  \\
\hline	
\end{tabular}}
\end{table}

% -- -- -- -- -- -- -- -- -- -- -- -- -- -- -- --
\section{Evaluation Results}

\subsection{RQ1: Performance of TRIAD} \label{sec:rq1-result}
Table \ref{tab:RQ1} presents the experiment results of five evaluated systems (rows).
For each system and each IR model (columns), we compared the performance of IR-ONLY, TAROT, LIA$_{rsc}$, and TRIAD. 
We also list the best result of LIA, denoted as LIA$_{best}$ based on VSM and LSI models for all datasets.
Sub-column 1 displays the average precision (AP), sub-column 2 shows the mean average precision (MAP), sub-column 3 shows the $p$-value of the F-measure significance test for TRIAD, and sub-column 4 shows Cliff’s $\delta$.
The best result of AP, MAP, and $p$-value $\leq$ 0.05 are in bold text for each approach.
When compared with IR-ONLY, TRIAD performs better in AP for 13 out of 15 cases\footnote{A ``case'' denotes a tuple of data set and IR model (i.e., VSM, LSI, and JS).} (from 0.9\% to 21\%, on average 10\%) and in MAP for 14 out of 15 cases (from 3\% to 35\%, on average 13\%).
The results show that TRIAD outperforms TAROT in AP for 13 out of 15 cases (from 4\% to 43\%, 18\% on average) and in MAP (from 6\% to 23\%, 13\% on average) for all 30 cases.
Compared with LIA$_{rcs}$, TRIAD performs better in AP for all 10 cases (from 5\% to 48\%, on average 26\%) and in MAP for 7 cases (from 6\% to 38\%, on average 14\%), and only performs worse in 3 cases (on average -6\%), specifically in WARC-VSM, LibEST-VSM, and LibEST-LSI.
Moreover, TRIAD even outperforms LIA$_{best}$ in AP for 8 out of 10 cases (from 0.8\% to 55\%, on average 18\%) and 4 out of 10 cases in MAP (from 2\% to 18\%, on average 8\%).

\begin{figure}[!bp]
\centering
% \vspace{1pt}
\subfigure[Dronology-VSM]{
	\label{fig5.1.1}
	\includegraphics[width=0.3\linewidth]{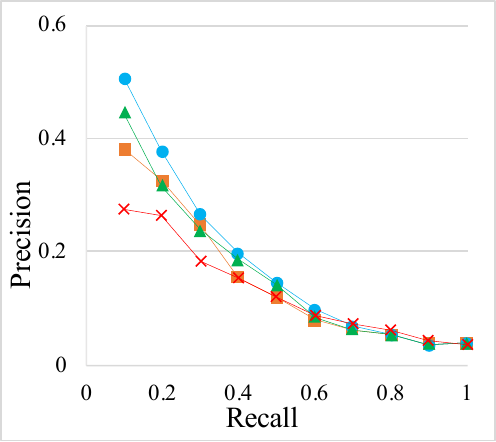}}
 \subfigure[Dronology-LSI]{
	\label{fig5.1.1}
	\includegraphics[width=0.3\linewidth]{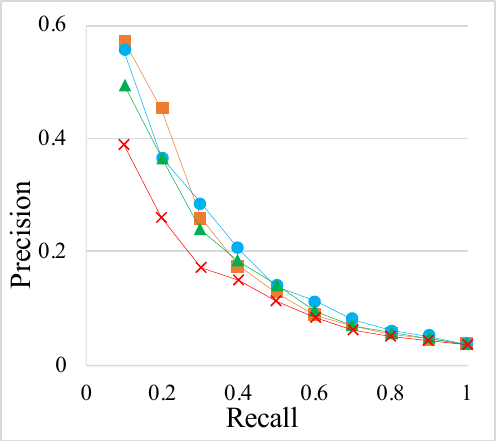}}
 \subfigure[Dronology-JSD]{
	\label{fig5.1.1}
	\includegraphics[width=0.3\linewidth]{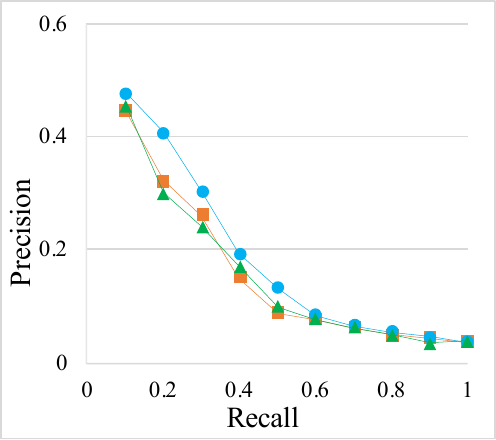}}
 
\subfigure[WARC-VSM]{
	\label{fig5.1.2}
	\includegraphics[width=0.3\linewidth]{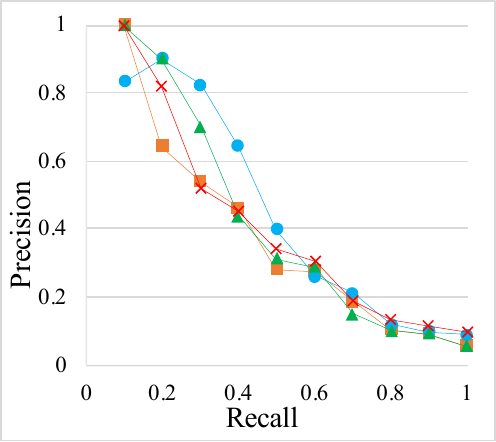}}
 \subfigure[WARC-LSI]{
	\label{fig5.1.2}
	\includegraphics[width=0.3\linewidth]{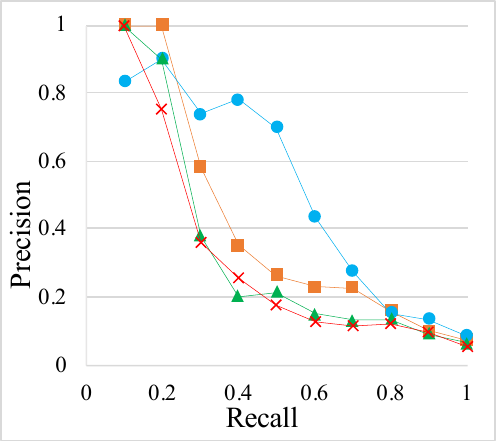}}
 \subfigure[WARC-JSD]{
	\label{fig5.1.2}
	\includegraphics[width=0.3\linewidth]{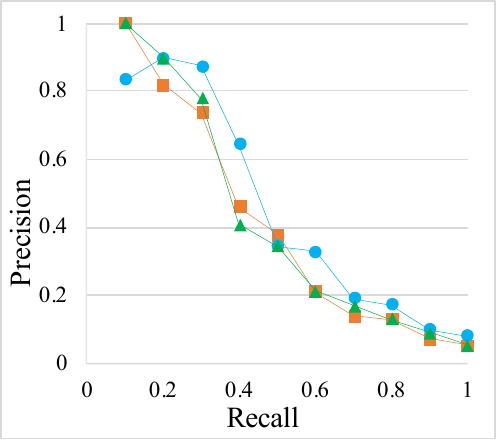}}
 
\subfigure[EasyClinic-VSM]{
	\label{fig5.1.3}
	\includegraphics[width=0.3\linewidth]{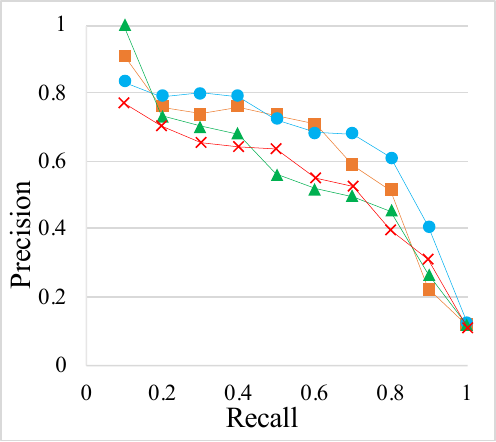}}
 \subfigure[EasyClinic-LSI]{
	\label{fig5.1.3}
	\includegraphics[width=0.3\linewidth]{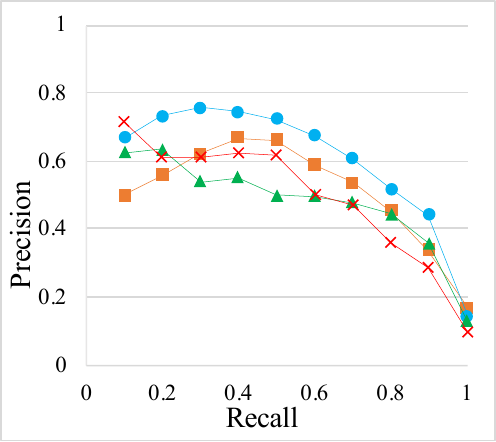}}
 \subfigure[EasyClinic-JSD]{
	\label{fig5.1.3}
	\includegraphics[width=0.3\linewidth]{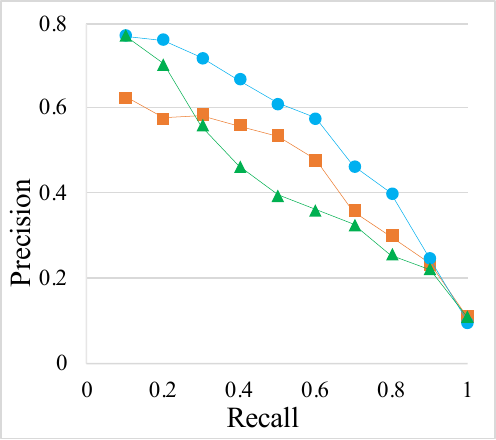}}
 
\subfigure[EBT-VSM]{
	\label{fig5.1.4}
	\includegraphics[width=0.3\linewidth]{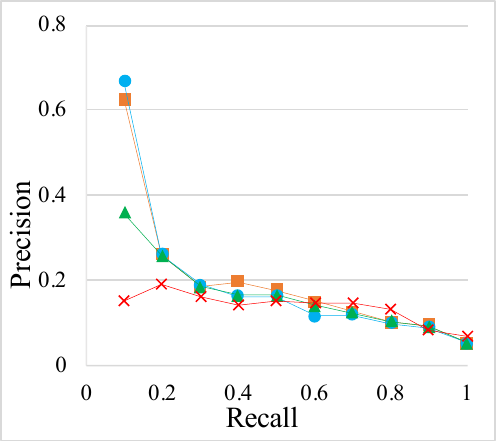}}
 \subfigure[EBT-LSI]{
	\label{fig5.1.4}
	\includegraphics[width=0.3\linewidth]{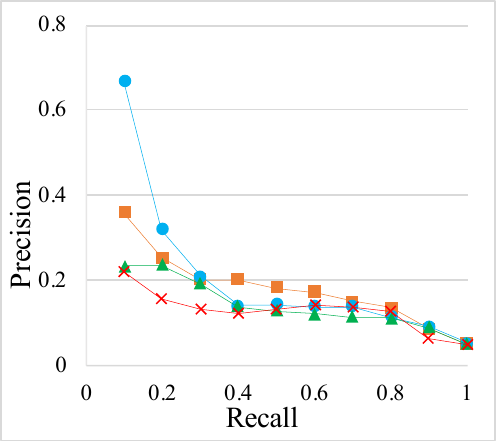}}
 \subfigure[EBT-JSD]{
	\label{fig5.1.4}
	\includegraphics[width=0.3\linewidth]{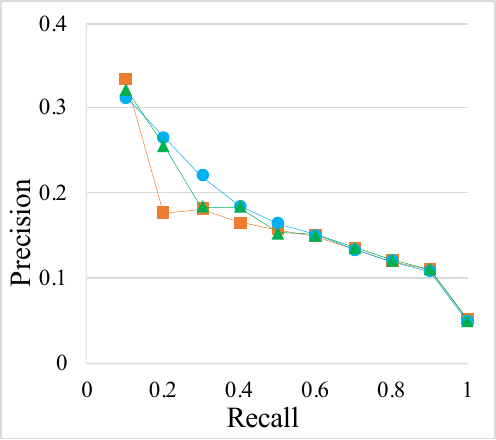}}
 
\subfigure[LibEST-VSM]{
	\label{fig5.1.5}
	\includegraphics[width=0.3\linewidth]{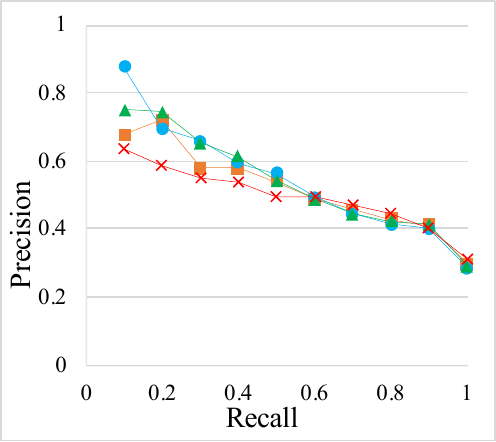}}
\subfigure[LibEST-LSI]{
	\label{fig5.1.5}
	\includegraphics[width=0.3\linewidth]{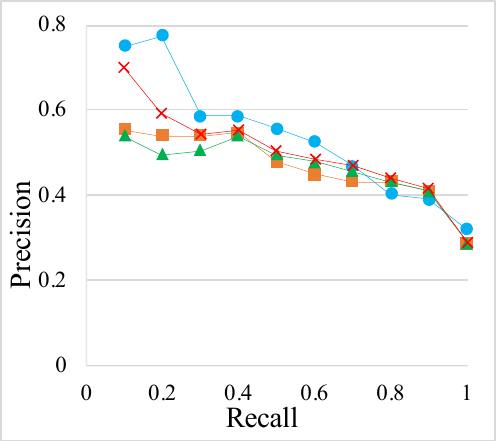}}
\subfigure[LibEST-JSD]{
	\label{fig5.1.5}
	\includegraphics[width=0.3\linewidth]{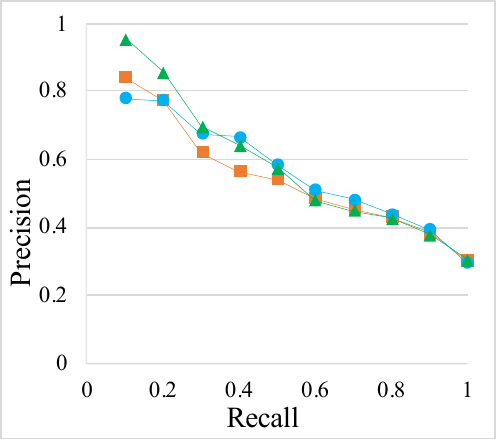}}

\includegraphics[width=0.8\linewidth]{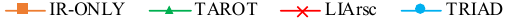} \\

\caption{Precision/Recall curves grouped by evaluated systems and IR models (VSM, LSI, and JS).}
\label{fig:pr}
\end{figure}

Notably, in 26 out of 40 cases (82.5\%), the F-measure for the result of TRIAD is significantly higher than three baseline approaches (p-value $\leq$ 0.05) at each level of recall, indicating that TRIAD outperforms baseline approaches in the majority of cases.
Figure \ref{fig:pr} illustrates precision-recall curves for the four approaches grouped by each system and IR model.
We now use the adapted excerpt from the Dronology system to demonstrate why TRIAD outperforms the baseline approaches.
For requirement RE-691 and two traces class \texttt{AFEmergencyComponent}, and \texttt{AFInfoBox} as shown in Figure~\ref{fig:approach-case}, the IR values calculated based on the VSM model between RE-691 and \texttt{AFInfoBox} and \texttt{AFEmergencyComponent} are 0.002 and 0, and ranks in RE-691's candidate trace list are 95/184 and 184/184, respectively. 
This indicates that there is a vast abstraction gap between RE-691 and these two classes, which causes them to share almost no semantic information.
Therefore, in this situation, IR-ONLY does not work anymore, and TAROT cannot extract any consensual biterms from these three artifacts.
For LIA$_{rcs}$, it uses direct links (i.e., from source to target directly) and transitive links in its combined strategies.
Hence, it is limited to only using transitive links to overcome this abstraction gap, which can promote their ranks to 59/184 and 120/184.
For TRIAD, it first extends RE-691 with intermediate-centric biterms from itself, including (select, uav) and (uav, oper).
Then it extends RE-691 with biterms from two highly related intermediate artifacts, such as (emerg, oper) and (appli, oper) from DD-694.
Furthermore, \texttt{AFEmergencyComponent} can be extended with biterms (select, uav) and (emerg, oper) from DD-694.
Therefore, TRIAD can overcome this abstraction gap with intermediate-centric biterms, which promote ranks to 10/184 and 3/184.
Furthermore, TRIAD can further use outer- and inner-transitive links, e.g., RE-691→DD-694→\texttt{AFEmergencyComponent} and RE-691→DD-694→\linebreak DD-647→\texttt{AFInfoBox} to overcome this abstraction gap.

When compared with COMET on EBT, TRIAD on EBT-JS and EBT-VSM had the lowest and highest APs, denoted as TRIAD$_{low}$ and TRIAD$_{high}$ in Table~\ref{tab:RQ1}, respectively. TRIAD$_{low}$ significantly outperforms all four cases on EBT, with improvements of 28\% in AP and 5\% in MAP compared to COMET$_{map}$, and 22\% in AP and 6\% in MAP compared to COMET$_{nuts}$.
Compared to COMET on LibEST, LibEST-LSI, and LibEST-JS are selected as TRIAD$_{low}$ and TRIAD$_{high}$, respectively.
For LibEST, COMET$_{nuts}$ and COMET$_{map}$ can both outperform TRIAD$_{low}$ in AP and MAP. 
However, TRIAD$_{high}$ can still outperform both COMET$_{nuts}$ and COMET$_{map}$ in MAP by 1.6\% and 1.3\%, respectively.
It is worth noting that COMET draws strength from multiple perspectives including tuned textual similarity, developer feedback, and transitive links, whereas TRIAD solely relies on textual similarities (where transitive links are deduced). But TRIAD can still outperform COMET in 3 out of 4 cases in both APs and MAPs on the EBT and LibEST datasets.

Furthermore, we noted that TRIAD performed worse in MAPs on two cases, i.e., LibEST-VSM and LibEST-LSI with a decrease of 4.96 (6.41\%) and 6.35 (8.34\%) compared to the performance of LIA$_{rcs}$, respectively.
Accordingly, we first found that the source and target artifacts of LibEST were enriched with an excessive number of biterms due to the numerous functions contained in both test code and source code files (all written in C). 
Particularly, the average number of biterms for each source artifact increased from 29 to 383 (with the largest number of 505), and for each target artifact, it increased from 190 to 551 (with the largest number of 1,517).
Another possible reason is that unlike other evaluated systems, the intermediate artifacts of LibEST are also code for testing that are semantically too far away from source targets (requirements) while too close to target artifact (source code), thus weakening TRIAD's performance.
Further discussions are in Section \ref{sec:threats}.

\subsection{RQ2: Ablation Study on TRIAD}
Table \ref{tab:RQ2} shows the experiment results of five evaluated systems (rows), where ``b'' denotes TRIAD using extended artifacts with intermediate-centric biterms only; ``o'' denotes TRIAD using outer-transitive links only, and ``i'' denotes TRIAD using inter-transitive links only.
For each system and each IR model (columns), sub-column 1 and sub-column 2 show the average precision (AP) and mean average precision (MAP), respectively.
For each evaluated system, the best results of AP and MAP on each IR model are in bold text.
We first compared the performance of each individual part of TRIAD (i.e., ``b'', ``i'', and ``o'').
From the table, we can observe that ``b'' outperforms IR-ONLY in 10 out of 15 cases on AP (improvement varies from 0.37\% to 23.39\%, 10.28\% on average) and in 13 out of 15 cases on MAP (from 1.02\% to 13.82\%, 6.72\% on average), which indicates that TRIAD can make improvement by extending artifacts with intermediate-centric consensual biterms. 
Then we compared TRIAD's performance with those using transitive links only.
We can observe that ``o'' outperforms IR-ONLY in 14 out of 15 cases on AP (improvement varies from 0.38\% to 13.48\%, 5.66\% on average) and in 13 out of 15 cases on MAP (from 0.61\% to 14.53\%, 6.38\% on average), which indicates that leveraging outer-transitive links achieve better performance.
Meanwhile, the results show that ``o+i'' outperform ``o'' in 10 out of 15 cases on AP (from 0.96\% to 6.62\%, 3.08\% on average), and 13 out of 15 cases on MAP (from 0.43\% to 5.73\%, 2.66\% on average), respectively.
This result indicates that using outer-inner combined transitive links can further improve the performance. 

In addition, the results show that ``b+o'' outperforms ``b'' in 10 out of 15 cases on AP (from 1.3\% to 10.41\%, 6.01\% on average), and 12 out of 15 cases on MAP (from 0.13\% to 11.06\%, 5.68\% on average) respectively, which means incorporating outer-transitive links can improve the performance compared to the situation when only intermediate-centric biterms are used.
Moreover, the results show that ``b+o+i'' outperform ``b+o'' in 9 out of 15 cases on AP (from 0.13\% to 3.53\%, 1.69\% on average), and 11 out of 15 cases on MAP (from 0.11\% to 7.09\%, 2.55\% on average), respectively.
This result indicates that combining outer- and inner-transitive links with intermediate-centric biterms can further improve the performance.
In conclusion, including all parts of TRIAD achieves the overall best performance.
Moreover, consensual biterms can complement transitive links to improve automated traceability recovery.

\begin{table}[!tb]
\caption{Performance of each part of TRIAD (RQ2) }
\renewcommand\arraystretch{1.0}
\resizebox{\linewidth}{!}{
\scriptsize
\label{tab:RQ2}
\begin{tabular} {l|l|ll|ll|ll}
\hline

&      & \multicolumn{2}{c}{VSM} & \multicolumn{2}{c}{LSI} & \multicolumn{2}{c}{JS} \\
\cline{3-4} 	\cline{5-6}	\cline{7-8} 
&      & AP & MAP    & AP & MAP       & AP & MAP   \\ 

\hline%\midrule		
\multirow{6}{*}{dronology}  & IR-ONLY & 18.93 & 35.88 & \textbf{22.74} & 40.75 & 17.63 & 34.38 \\
                            & b       & 18.67 & 37.44 & 19.21 & 44.80 & 18.29 & 39.50 \\
                            & o       & 19.70 & 38.05 & 22.39 & 41.00 & 19.72 & 38.54 \\
                            & b+o     & 19.59 & 41.40 & 20.63 & 47.73 & 19.85 & 43.32 \\
                            & o+i     & \textbf{20.08} & 39.85 & 22.04 & 41.74 & 19.91 & 40.75 \\
                            & b+o+i   & 20.07 & \textbf{43.25} & 21.13 & \textbf{49.07} & \textbf{20.55} & \textbf{46.39} \\
                            \hline
\multirow{6}{*}{warc}       & IR-ONLY & 41.41 & 65.01 & 44.20  & 61.24 & 43.61 & 66.63 \\
                            & b       & \textbf{48.36} & 70.49 & \textbf{54.54} & 63.22 & \textbf{47.45} & 70.82 \\
                            & o       & 42.84 & 66.05 & 47.64 & 69.65 & 46.13 & 67.64 \\
                            & b+o     & 46.12 & 70.49 & 51.45 & 63.30 & 45.75 & 71.13 \\
                            & o+i     & 45.12 & 67.74 & 48.94 & \textbf{72.35} & 45.45 & 68.10 \\
                            & b+o+i   & 45.74 & \textbf{70.49} & 51.26 & 63.77 & 46.21 & \textbf{71.21} \\
                            \hline
\multirow{6}{*}{easyclinic} & IR-ONLY & 65.39 & 76.51 & 53.51 & 69.47 & 46.75 & 62.67 \\
                            & b       & \textbf{70.21} & 78.10  & 65.87 & 74.32 & 54.56 & 65.01 \\
                            & o       & 65.64 & 76.17 & 55.65 & 71.11 & 53.05 & 67.04 \\
                            & b+o     & 69.80 & \textbf{80.04} & \textbf{65.20} & \textbf{78.06} & 55.95 & 68.37 \\
                            & o+i     & 64.37 & 75.31 & 58.47 & 71.79 & 53.77 & 66.81 \\
                            & b+o+i   & 68.32 & 79.05 & 64.55 & 78.01 & \textbf{56.35} & \textbf{69.10} \\
                            \hline
\multirow{6}{*}{ebt}        & IR-ONLY & 23.25 & 33.03 & 20.33 & 39.44 & 20.81 & 33.39 \\
                            & b       & 21.67 & 35.60 & 19.88 & 41.33 & 20.85 & 33.73 \\
                            & o       & 24.22 & 37.83 & 22.51 & 41.28 & 21.85 & 35.51 \\
                            & b+o     & 23.01 & 37.96 & 21.95 & 40.65 & \textbf{22.73} & 37.46 \\
                            & o+i     & \textbf{24.78} & \textbf{38.27} & \textbf{24.00} & \textbf{42.10} & 20.08 & 37.36 \\
                            & b+o+i   & 23.04 & 38.10 & 22.30 & 40.55 & 21.00 & \textbf{39.68} \\
                            \hline
\multirow{6}{*}{libest}     & IR-ONLY & 55.25 & 73.30 & 50.94 & 62.54 & 57.69 & 66.63 \\
                            & b       & 54.41 & 71.41 & 51.13 & 61.83 & 59.03 & \textbf{75.84} \\
                            & o       & 56.56 & 73.27 & 53.37 & 68.06 & 58.54 & 69.05 \\
                            & b+o     & 56.18 & 72.08 & 54.46 & 67.67 & \textbf{59.80} & 73.89 \\
                            & o+i     & \textbf{57.33} & \textbf{74.04} & 55.06 & \textbf{71.01} & 57.72 & 69.35 \\
                            & b+o+i   & 56.86 & 72.40 & \textbf{55.62} & 69.81 & 59.13 & 75.26 \\   
\hline

\end{tabular}}
\end{table}

% -- -- -- -- -- -- -- -- -- -- -- -- -- -- -- --
\section{Discussion and Threats to validity} \label{sec:threats}

\vspace{2mm} \noindent
\textbf{Implications for Researchers.}
As we discussed at the end of Section \ref{sec:rq1-result}, TRIAD performed worse in MAPs on two experiment cases: LibEST-VSM and LibEST-LSI. This result is caused by two situations: (1) the source and target artifacts of LibEST were enriched with an excessive number of biterms due to the large size code files written in C; (2) the intermediate artifact of LibEST is test code and it is not sufficient to bridge the abstraction gap between source (requirements) and target (source code) artifacts. The former situation implies that when tracing a large-size code file, it is likely to implement multiple requirements, and requires fine-granular analysis on methods instead of code files (or classes in Java). The latter situation implies that when introducing intermediate artifacts to improve source-to-target tracing, relatively ``middle-level'' artifacts are more suitable to bridge the abstraction gap, e.g., the design definitions in Dronology, rather than the test code in LibEST.

\vspace{2mm} \noindent
\textbf{Implications for Practitioners.}
While the results are promising and performance keeps improving, so far the practitioners can still not fully rely on automated tracing approaches, but we obtain insights into why that is and hence motivate further investigation into better bridging the abstraction gap. Furthermore, to fine-tune TRIAD for optimized performance in practice when necessary, users can first modestly tune up threshold $m$ (0.5 by default) when the quantity and quality of artifact texts are better, and vice versa. Second, if users can effectively estimate the number of relevant traces, we suggest users tune up threshold $t$ (3 by default) when the system tends to have a larger number of relevant traces.

\vspace{2mm} \noindent
\textbf{Special Circumstances.} Our approach is likely to encounter the following two special cases when applied to a system: (1) the system only contains source and target artifacts with no available intermediate artifacts; (2) the system already contains existing traces that are either source-to-intermediate or intermediate-to-target links.
For the first case, TRIAD can still leverage inner-transitive relationships within source artifacts to adjust trace scores between source and target artifacts.
However, in this case we think that the performance of TRIAD is likely to decrease because it cannot leverage the implicit semantics from the missing intermediate artifacts.
For the other case, when the system has existing high-quality (i.e., correct and complete), source-to-intermediate or intermediate-to-target traces, TRIAD can directly leverage these existing traces and we expect TRIAD to achieve better performance because the originally deduced transitive links may be erroneous and incomplete.
Furthermore, when the quality of existing traces cannot be guaranteed, it would be interesting to consider both the existing traces and the deduced traces in TRIAD and try to achieve the overall best performance. 
% How can TRIAD work better under the discussed special circumstances is in future work.
The exploration of how TRIAD can enhance its effectiveness under the discussed specific circumstances is in future work.

\vspace{2mm} \noindent
\textbf{Threats to validity.}
To avoid any bias or fine-tuning, we used the same threshold $m$ and $t$ in all five evaluated systems on each of the three IR models.
We use the officially published code that is online available \cite{conf/icsm/RodriguezCF21} to replicate LIA as our baseline approach.
For COMET, to avoid any bias by consulting existing research \cite{Hey-icsme2021}, we directly use the result lists that are provided in its replication package ~\cite{DBLP:conf/icse/MoranPBMPSJ20} to compute the final experiment metrics.
Meanwhile, we also use the same pre-processed data of COMET, which is also available in the replication package, as the input for TRIAD to ensure a fair comparison in our evaluation.
Furthermore, we choose TRIAD with the lowest and highest AP performances across the three IR models to compare with COMET because the authors that proposed COMET only use AP in their work.
Another possible internal threat to our approach is that we cannot guarantee 100\% accuracy in segmenting texts, tagging POSs, and parsing dependencies based on Stanford CoreNLP. 
However, existing work has reported that the accuracy of off-the-shelf NLP tools is acceptable when analyzing texts in the context of proper sentences and grammatical structures \cite{DBLP:journals/infsof/AliCHH19}, and we found no obvious errors in the output of Stanford CoreNLP as well.

% -- -- -- -- -- -- -- -- -- -- -- -- -- -- -- --
\section{Conclusion}
\label{sec:conclusion}

In this work, we addressed the abstraction gap problem between artifacts to be traced. We proposed the use of consensual biterms extraction to further deduce transitive links between both the source and intermediate artifacts, and the intermediate and target artifacts. The goal is to bridge the abstraction gap better to improve automated tracing. 
This novel use of transitive links and consensual biterms was implemented in our approach \textbf{TRIAD} (\textbf{T}raceability \textbf{R}ecovery 
by b\textbf{I}term-enh\textbf{A}nced \textbf{D}eduction of transitive links).

To assess how TRIAD tackles the traceability problem of filling the abstraction gap, we conduct an empirical evaluation based on five systems widely used in literature. TRIAD was compared with four state-of-the-art approaches based on metrics of average precision and mean average Precision, along with significance tests on F-Measure. 
Results show that TRIAD outperforms the four baselines in most of the cases. We also observed that using outer-inner combined transitive links can further alleviate the abstraction gap problem. 
Future work will primarily focus on better bridging the abstraction gap by more sophisticated identification of inner- and outer transitive links.

% -- -- -- -- -- -- -- -- -- -- -- -- -- -- -- --
\section{Data availability}
Our replication package with the TRIAD source code and datasets used in the evaluation is available online~\cite{TRIADcode, TRIADdataset}.

% -- -- -- -- -- -- -- -- -- -- -- -- -- -- -- --
\section{Acknowledgements}
This work is supported by the National Natural Science Foundation of China (No.62025202, No.72371125, No.62072227, No.62202219, and No.62302210), the China Scholarship Council (No.202206190172), the Innovation Project and Overseas Open Project of State Key Laboratory for Novel Software Technology (Nanjing University) (ZZKT2022A25, KFKT2022A09, KFKT2023A09, KFKT2023A10), and \linebreak the Jiangsu Provincial Key Research and Development Program (No.BE2021002-2).
We also gratefully acknowledge the Austrian Science Fund (FWF), grand No. P31989 and No.P34805-N.

\bibliographystyle{ACM-Reference-Format}
\bibliography{references}


\begin{thebibliography}{72}

%%% the Web address.

\ifx \showCODEN    \undefined \def \showCODEN     #1{\unskip}     \fi
\ifx \showDOI      \undefined \def \showDOI       #1{#1}\fi
\ifx \showISBNx    \undefined \def \showISBNx     #1{\unskip}     \fi
\ifx \showISBNxiii \undefined \def \showISBNxiii  #1{\unskip}     \fi
\ifx \showISSN     \undefined \def \showISSN      #1{\unskip}     \fi
\ifx \showLCCN     \undefined \def \showLCCN      #1{\unskip}     \fi
\ifx \shownote     \undefined \def \shownote      #1{#1}          \fi
\ifx \showarticletitle \undefined \def \showarticletitle #1{#1}   \fi
\ifx \showURL      \undefined \def \showURL       {\relax}        \fi
% invisible to TeX
\providecommand\bibfield[2]{#2}
\providecommand\bibinfo[2]{#2}
\providecommand\natexlab[1]{#1}
\providecommand\showeprint[2][]{arXiv:#2}

\bibitem[CoE(2023a)]%
        {CoEST}
 \bibinfo{year}{2023}\natexlab{a}.
\newblock \bibinfo{title}{Center of Excellence for Software and Systems
  Traceability}.
\newblock \bibinfo{howpublished}{\url{http://www.coest.org/}}.
\newblock


\bibitem[CoE(2023b)]%
        {CoESTDataset}
 \bibinfo{year}{2023}\natexlab{b}.
\newblock \bibinfo{title}{CoEST community datasets}.
\newblock
  \bibinfo{howpublished}{\url{http://sarec.nd.edu/coest/datasets.html}}.
\newblock


\bibitem[Lib(2023)]%
        {LibESTDataset}
 \bibinfo{year}{2023}\natexlab{}.
\newblock \bibinfo{title}{Comet Data Replication Package: {LibEST}}.
\newblock
  \bibinfo{howpublished}{\url{https://gitlab.com/SEMERU-Code-Public/Data/icse20-comet-data-replication-package/-/tree/main/LibEST}}.
\newblock


\bibitem[dro(2023)]%
        {dronologyDataset}
 \bibinfo{year}{2023}\natexlab{}.
\newblock \bibinfo{title}{Dronology Datasets}.
\newblock \bibinfo{howpublished}{\url{https://dronology.info/datasets/}}.
\newblock


\bibitem[src(2023)]%
        {srcML}
 \bibinfo{year}{2023}\natexlab{}.
\newblock \bibinfo{title}{srcML}.
\newblock \bibinfo{howpublished}{\url{https://www.srcml.org/}}.
\newblock


\bibitem[TRI(2023a)]%
        {TRIADcode}
 \bibinfo{year}{2023}\natexlab{a}.
\newblock \bibinfo{title}{TRIAD code}.
\newblock \bibinfo{howpublished}{\url{https://github.com/huiAlex/TRIAD}}.
\newblock


\bibitem[TRI(2023b)]%
        {TRIADdataset}
 \bibinfo{year}{2023}\natexlab{b}.
\newblock \bibinfo{title}{TRIAD dataset}.
\newblock
  \bibinfo{howpublished}{\url{https://doi.org/10.5281/zenodo.10430771}}.
\newblock


\bibitem[Abadi et~al\mbox{.}(2008)]%
        {DBLP:conf/iwpc/AbadiNS08}
\bibfield{author}{\bibinfo{person}{Ahron Abadi}, \bibinfo{person}{Mordechai
  Nisenson}, {and} \bibinfo{person}{Yahalomit Simionovici}.}
  \bibinfo{year}{2008}\natexlab{}.
\newblock \showarticletitle{A Traceability Technique for Specifications}. In
  \bibinfo{booktitle}{\emph{16th {IEEE} International Conference on Program
  Comprehension}}. \bibinfo{publisher}{{IEEE}}, \bibinfo{pages}{103--112}.
\newblock


\bibitem[Ali et~al\mbox{.}(2019)]%
        {DBLP:journals/infsof/AliCHH19}
\bibfield{author}{\bibinfo{person}{Nasir Ali}, \bibinfo{person}{Haipeng Cai},
  \bibinfo{person}{Abdelwahab Hamou{-}Lhadj}, {and}
  \bibinfo{person}{Jameleddine Hassine}.} \bibinfo{year}{2019}\natexlab{}.
\newblock \showarticletitle{Exploiting Parts-of-Speech for effective automated
  requirements traceability}.
\newblock \bibinfo{journal}{\emph{Inf. Softw. Technol.}}  \bibinfo{volume}{106}
  (\bibinfo{year}{2019}), \bibinfo{pages}{126--141}.
\newblock
\urldef\tempurl%
\url{https://doi.org/10.1016/j.infsof.2018.09.009}
\showDOI{\tempurl}


\bibitem[Antoniol et~al\mbox{.}(2002)]%
        {DBLP:journals/tse/AntoniolCCLM02}
\bibfield{author}{\bibinfo{person}{Giuliano Antoniol}, \bibinfo{person}{Gerardo
  Canfora}, \bibinfo{person}{Gerardo Casazza}, \bibinfo{person}{Andrea~De
  Lucia}, {and} \bibinfo{person}{Ettore Merlo}.}
  \bibinfo{year}{2002}\natexlab{}.
\newblock \showarticletitle{Recovering Traceability Links between Code and
  Documentation}.
\newblock \bibinfo{journal}{\emph{{IEEE} Trans. Software Eng.}}
  \bibinfo{volume}{28}, \bibinfo{number}{10} (\bibinfo{year}{2002}),
  \bibinfo{pages}{970--983}.
\newblock


\bibitem[Baezayates and Ribeironeto(2011)]%
        {Baezayates2004Modern}
\bibfield{author}{\bibinfo{person}{Ricardo Baezayates} {and}
  \bibinfo{person}{Berthier Ribeironeto}.} \bibinfo{year}{2011}\natexlab{}.
\newblock \bibinfo{booktitle}{\emph{Modern information retrieval}}.
\newblock \bibinfo{publisher}{Addison-Wesley Publishing CompanyUnited States}.
\newblock


\bibitem[Bassett and Deride(2019)]%
        {BassettD19}
\bibfield{author}{\bibinfo{person}{Robert Bassett} {and} \bibinfo{person}{Julio
  Deride}.} \bibinfo{year}{2019}\natexlab{}.
\newblock \showarticletitle{Maximum a posteriori estimators as a limit of Bayes
  estimators}.
\newblock \bibinfo{journal}{\emph{Math. Program.}} \bibinfo{volume}{174},
  \bibinfo{number}{1-2} (\bibinfo{year}{2019}), \bibinfo{pages}{129--144}.
\newblock
\urldef\tempurl%
\url{https://doi.org/10.1007/S10107-018-1241-0}
\showDOI{\tempurl}


\bibitem[Biggerstaff et~al\mbox{.}(1993)]%
        {conf/icse/BiggerstaffMW93}
\bibfield{author}{\bibinfo{person}{Ted~J. Biggerstaff},
  \bibinfo{person}{Bharat~G. Mitbander}, {and} \bibinfo{person}{Dallas~E.
  Webster}.} \bibinfo{year}{1993}\natexlab{}.
\newblock \showarticletitle{The Concept Assignment Problem in Program
  Understanding}. In \bibinfo{booktitle}{\emph{15th International Conference on
  Software Engineering}}, \bibfield{editor}{\bibinfo{person}{Victor~R. Basili},
  \bibinfo{person}{Richard~A. DeMillo}, {and} \bibinfo{person}{Takuya
  Katayama}} (Eds.). \bibinfo{publisher}{{IEEE}/{ACM}},
  \bibinfo{pages}{482--498}.
\newblock


\bibitem[Cheng et~al\mbox{.}(2014)]%
        {6778764BTM}
\bibfield{author}{\bibinfo{person}{Xueqi Cheng}, \bibinfo{person}{Xiaohui Yan},
  \bibinfo{person}{Yanyan Lan}, {and} \bibinfo{person}{Jiafeng Guo}.}
  \bibinfo{year}{2014}\natexlab{}.
\newblock \showarticletitle{BTM: Topic Modeling over Short Texts}.
\newblock \bibinfo{journal}{\emph{IEEE Transactions on Knowledge and Data
  Engineering}} \bibinfo{volume}{26}, \bibinfo{number}{12}
  (\bibinfo{year}{2014}), \bibinfo{pages}{2928--2941}.
\newblock
\urldef\tempurl%
\url{https://doi.org/10.1109/TKDE.2014.2313872}
\showDOI{\tempurl}


\bibitem[Chikofsky and II(1990)]%
        {journals/software/ChikofskyC90}
\bibfield{author}{\bibinfo{person}{Elliot~J. Chikofsky} {and}
  \bibinfo{person}{James H.~Cross II}.} \bibinfo{year}{1990}\natexlab{}.
\newblock \showarticletitle{Reverse Engineering and Design Recovery: {A}
  Taxonomy}.
\newblock \bibinfo{journal}{\emph{{IEEE} Softw.}} \bibinfo{volume}{7},
  \bibinfo{number}{1} (\bibinfo{year}{1990}), \bibinfo{pages}{13--17}.
\newblock
\urldef\tempurl%
\url{https://doi.org/10.1109/52.43044}
\showDOI{\tempurl}


\bibitem[Cleland{-}Huang et~al\mbox{.}(2014)]%
        {DBLP:conf/icse/Cleland-HuangGHMZ14}
\bibfield{author}{\bibinfo{person}{Jane Cleland{-}Huang},
  \bibinfo{person}{Orlena Gotel}, \bibinfo{person}{Jane~Huffman Hayes},
  \bibinfo{person}{Patrick M{\"{a}}der}, {and} \bibinfo{person}{Andrea
  Zisman}.} \bibinfo{year}{2014}\natexlab{}.
\newblock \showarticletitle{Software traceability: trends and future
  directions}. In \bibinfo{booktitle}{\emph{Future of Software Engineering}}.
  \bibinfo{publisher}{{ACM}}, \bibinfo{pages}{55--69}.
\newblock


\bibitem[Cleland{-}Huang et~al\mbox{.}(2005)]%
        {DBLP:conf/re/Cleland-HuangSDZ05}
\bibfield{author}{\bibinfo{person}{Jane Cleland{-}Huang},
  \bibinfo{person}{Raffaella Settimi}, \bibinfo{person}{Chuan Duan}, {and}
  \bibinfo{person}{Xuchang Zou}.} \bibinfo{year}{2005}\natexlab{}.
\newblock \showarticletitle{Utilizing Supporting Evidence to Improve Dynamic
  Requirements Traceability}. In \bibinfo{booktitle}{\emph{13th {IEEE}
  International Conference on Requirements Engineering}}.
  \bibinfo{publisher}{{IEEE}}, \bibinfo{pages}{135--144}.
\newblock


\bibitem[Cleland{-}Huang et~al\mbox{.}(2018)]%
        {DBLP:conf/icse/Cleland-HuangVB18}
\bibfield{author}{\bibinfo{person}{Jane Cleland{-}Huang},
  \bibinfo{person}{Michael Vierhauser}, {and} \bibinfo{person}{Sean Bayley}.}
  \bibinfo{year}{2018}\natexlab{}.
\newblock \showarticletitle{Dronology: an incubator for cyber-physical systems
  research}. In \bibinfo{booktitle}{\emph{40th International Conference on
  Software Engineering}}. \bibinfo{publisher}{{ACM}},
  \bibinfo{pages}{109--112}.
\newblock
\urldef\tempurl%
\url{https://doi.org/10.1145/3183399.3183408}
\showDOI{\tempurl}


\bibitem[De{ }Lucia et~al\mbox{.}(2012)]%
        {DBLP:books/daglib/p/LuciaMOP12}
\bibfield{author}{\bibinfo{person}{Andrea De{ }Lucia}, \bibinfo{person}{Andrian
  Marcus}, \bibinfo{person}{Rocco Oliveto}, {and} \bibinfo{person}{Denys
  Poshyvanyk}.} \bibinfo{year}{2012}\natexlab{}.
\newblock \showarticletitle{Information Retrieval Methods for Automated
  Traceability Recovery}.
\newblock In \bibinfo{booktitle}{\emph{Software and Systems Traceability}},
  \bibfield{editor}{\bibinfo{person}{Jane Cleland{-}Huang},
  \bibinfo{person}{Olly Gotel}, {and} \bibinfo{person}{Andrea Zisman}} (Eds.).
  \bibinfo{publisher}{Springer}, \bibinfo{pages}{71--98}.
\newblock


\bibitem[De{ }Lucia et~al\mbox{.}(2006)]%
        {DBLP:conf/icsm/LuciaOS06}
\bibfield{author}{\bibinfo{person}{Andrea De{ }Lucia}, \bibinfo{person}{Rocco
  Oliveto}, {and} \bibinfo{person}{Paola Sgueglia}.}
  \bibinfo{year}{2006}\natexlab{}.
\newblock \showarticletitle{Incremental Approach and User Feedbacks: a Silver
  Bullet for Traceability Recovery}. In \bibinfo{booktitle}{\emph{22nd {IEEE}
  International Conference on Software Maintenance}}.
  \bibinfo{publisher}{{IEEE}}, \bibinfo{pages}{299--309}.
\newblock


\bibitem[De{ }Lucia et~al\mbox{.}(2011)]%
        {DBLP:conf/iwpc/LuciaPOPP11}
\bibfield{author}{\bibinfo{person}{Andrea De{ }Lucia},
  \bibinfo{person}{Massimiliano~Di Penta}, \bibinfo{person}{Rocco Oliveto},
  \bibinfo{person}{Annibale Panichella}, {and} \bibinfo{person}{Sebastiano
  Panichella}.} \bibinfo{year}{2011}\natexlab{}.
\newblock \showarticletitle{Improving {IR}-based Traceability Recovery Using
  Smoothing Filters}. In \bibinfo{booktitle}{\emph{19th {IEEE} International
  Conference on Program Comprehension}}. \bibinfo{publisher}{{IEEE}},
  \bibinfo{pages}{21--30}.
\newblock


\bibitem[Diaz et~al\mbox{.}(2013)]%
        {DBLP:conf/iwpc/DiazBMOTL13}
\bibfield{author}{\bibinfo{person}{Diana Diaz}, \bibinfo{person}{Gabriele
  Bavota}, \bibinfo{person}{Andrian Marcus}, \bibinfo{person}{Rocco Oliveto},
  \bibinfo{person}{Silvia Takahashi}, {and} \bibinfo{person}{Andrea~De Lucia}.}
  \bibinfo{year}{2013}\natexlab{}.
\newblock \showarticletitle{Using code ownership to improve IR-based
  Traceability Link Recovery}. In \bibinfo{booktitle}{\emph{21st {IEEE}
  International Conference on Program Comprehension}}.
  \bibinfo{publisher}{{IEEE} Computer Society}, \bibinfo{pages}{123--132}.
\newblock
\urldef\tempurl%
\url{https://doi.org/10.1109/ICPC.2013.6613840}
\showDOI{\tempurl}


\bibitem[Dit et~al\mbox{.}(2013)]%
        {DBLP:journals/ese/DitRP13}
\bibfield{author}{\bibinfo{person}{Bogdan Dit}, \bibinfo{person}{Meghan
  Revelle}, {and} \bibinfo{person}{Denys Poshyvanyk}.}
  \bibinfo{year}{2013}\natexlab{}.
\newblock \showarticletitle{Integrating information retrieval, execution and
  link analysis algorithms to improve feature location in software}.
\newblock \bibinfo{journal}{\emph{Empirical Software Engineering}}
  \bibinfo{volume}{18}, \bibinfo{number}{2} (\bibinfo{year}{2013}),
  \bibinfo{pages}{277--309}.
\newblock


\bibitem[Dong et~al\mbox{.}(2022)]%
        {DBLP:conf/sigsoft/0001ZLWK22}
\bibfield{author}{\bibinfo{person}{Liming Dong}, \bibinfo{person}{He Zhang},
  \bibinfo{person}{Wei Liu}, \bibinfo{person}{Zhiluo Weng}, {and}
  \bibinfo{person}{Hongyu Kuang}.} \bibinfo{year}{2022}\natexlab{}.
\newblock \showarticletitle{Semi-supervised pre-processing for learning-based
  traceability framework on real-world software projects}. In
  \bibinfo{booktitle}{\emph{30th {ACM} Joint European Software Engineering
  Conference and Symposium on the Foundations of Software Engineering}},
  \bibfield{editor}{\bibinfo{person}{Abhik Roychoudhury},
  \bibinfo{person}{Cristian Cadar}, {and} \bibinfo{person}{Miryung Kim}}
  (Eds.). \bibinfo{publisher}{{ACM}}, \bibinfo{pages}{570--582}.
\newblock
\urldef\tempurl%
\url{https://doi.org/10.1145/3540250.3549151}
\showDOI{\tempurl}


\bibitem[Egyed et~al\mbox{.}(2010)]%
        {DBLP:conf/re/EgyedGG10}
\bibfield{author}{\bibinfo{person}{Alexander Egyed}, \bibinfo{person}{Florian
  Graf}, {and} \bibinfo{person}{Paul Gr{\"{u}}nbacher}.}
  \bibinfo{year}{2010}\natexlab{}.
\newblock \showarticletitle{Effort and Quality of Recovering
  Requirements-to-Code Traces: Two Exploratory Experiments}. In
  \bibinfo{booktitle}{\emph{18th {IEEE} International Requirements Engineering
  Conference}}. \bibinfo{publisher}{{IEEE}}, \bibinfo{pages}{221--230}.
\newblock


\bibitem[Falessi et~al\mbox{.}(2020)]%
        {journals/tse/FalessiRGC20}
\bibfield{author}{\bibinfo{person}{Davide Falessi}, \bibinfo{person}{Justin
  Roll}, \bibinfo{person}{Jin L.~C. Guo}, {and} \bibinfo{person}{Jane
  Cleland{-}Huang}.} \bibinfo{year}{2020}\natexlab{}.
\newblock \showarticletitle{Leveraging Historical Associations between
  Requirements and Source Code to Identify Impacted Classes}.
\newblock \bibinfo{journal}{\emph{{IEEE} Trans. Software Eng.}}
  \bibinfo{volume}{46}, \bibinfo{number}{4} (\bibinfo{year}{2020}),
  \bibinfo{pages}{420--441}.
\newblock
\urldef\tempurl%
\url{https://doi.org/10.1109/TSE.2018.2861735}
\showDOI{\tempurl}


\bibitem[Gao et~al\mbox{.}(2022a)]%
        {DBLP:journals/ese/GaoKMHLME22}
\bibfield{author}{\bibinfo{person}{Hui Gao}, \bibinfo{person}{Hongyu Kuang},
  \bibinfo{person}{Xiaoxing Ma}, \bibinfo{person}{Hao Hu},
  \bibinfo{person}{Jian L{\"{u}}}, \bibinfo{person}{Patrick M{\"{a}}der}, {and}
  \bibinfo{person}{Alexander Egyed}.} \bibinfo{year}{2022}\natexlab{a}.
\newblock \showarticletitle{Propagating frugal user feedback through closeness
  of code dependencies to improve IR-based traceability recovery}.
\newblock \bibinfo{journal}{\emph{Empir. Softw. Eng.}} \bibinfo{volume}{27},
  \bibinfo{number}{2} (\bibinfo{year}{2022}), \bibinfo{pages}{41}.
\newblock
\urldef\tempurl%
\url{https://doi.org/10.1007/s10664-021-10091-5}
\showDOI{\tempurl}


\bibitem[Gao et~al\mbox{.}(2022b)]%
        {Hui2022ASE}
\bibfield{author}{\bibinfo{person}{Hui Gao}, \bibinfo{person}{Hongyu Kuang},
  \bibinfo{person}{Kexin Sun}, \bibinfo{person}{Xiaoxing Ma},
  \bibinfo{person}{Alexander Egyed}, \bibinfo{person}{Patrick M\"{a}der},
  \bibinfo{person}{Guoping Rong}, \bibinfo{person}{Dong Shao}, {and}
  \bibinfo{person}{He Zhang}.} \bibinfo{year}{2022}\natexlab{b}.
\newblock \showarticletitle{Using Consensual Biterms from Text Structures of
  Requirements and Code to Improve IR-Based Traceability Recovery}. In
  \bibinfo{booktitle}{\emph{37th IEEE/ACM International Conference on Automated
  Software Engineering}} \emph{(\bibinfo{series}{ASE '22})}.
  \bibinfo{publisher}{ACM}, Article \bibinfo{articleno}{114}.
\newblock
\showISBNx{9781450394758}
\urldef\tempurl%
\url{https://doi.org/10.1145/3551349.3556948}
\showDOI{\tempurl}


\bibitem[Gethers et~al\mbox{.}(2011)]%
        {DBLP:conf/icsm/GethersOPL11}
\bibfield{author}{\bibinfo{person}{Malcom Gethers}, \bibinfo{person}{Rocco
  Oliveto}, \bibinfo{person}{Denys Poshyvanyk}, {and}
  \bibinfo{person}{Andrea~De Lucia}.} \bibinfo{year}{2011}\natexlab{}.
\newblock \showarticletitle{On integrating orthogonal information retrieval
  methods to improve traceability recovery}. In
  \bibinfo{booktitle}{\emph{{IEEE} 27th International Conference on Software
  Maintenance}}. \bibinfo{publisher}{{IEEE} Computer Society},
  \bibinfo{pages}{133--142}.
\newblock


\bibitem[Grundy et~al\mbox{.}(1998)]%
        {Grundy1998}
\bibfield{author}{\bibinfo{person}{J. Grundy}, \bibinfo{person}{J. Hosking},
  {and} \bibinfo{person}{W.B. Mugridge}.} \bibinfo{year}{1998}\natexlab{}.
\newblock \showarticletitle{Inconsistency management for multiple-view software
  development environments}.
\newblock \bibinfo{journal}{\emph{{IEEE} Transactions on Software Engineering}}
  \bibinfo{volume}{24}, \bibinfo{number}{11} (\bibinfo{year}{1998}),
  \bibinfo{pages}{960--981}.
\newblock
\urldef\tempurl%
\url{https://doi.org/10.1109/32.730545}
\showDOI{\tempurl}


\bibitem[Guo et~al\mbox{.}(2017)]%
        {DBLP:conf/icse/0004CC17}
\bibfield{author}{\bibinfo{person}{Jin Guo}, \bibinfo{person}{Jinghui Cheng},
  {and} \bibinfo{person}{Jane Cleland{-}Huang}.}
  \bibinfo{year}{2017}\natexlab{}.
\newblock \showarticletitle{Semantically enhanced software traceability using
  deep learning techniques}. In \bibinfo{booktitle}{\emph{39th International
  Conference on Software Engineering}},
  \bibfield{editor}{\bibinfo{person}{Sebasti{\'{a}}n Uchitel},
  \bibinfo{person}{Alessandro Orso}, {and} \bibinfo{person}{Martin~P.
  Robillard}} (Eds.). \bibinfo{publisher}{{IEEE}/{ACM}},
  \bibinfo{pages}{3--14}.
\newblock


\bibitem[Hadi and Fard(2020)]%
        {9240699}
\bibfield{author}{\bibinfo{person}{Mohammad~Abdul Hadi} {and}
  \bibinfo{person}{Fatemeh~H Fard}.} \bibinfo{year}{2020}\natexlab{}.
\newblock \showarticletitle{AOBTM: Adaptive Online Biterm Topic Modeling for
  Version Sensitive Short-texts Analysis}. In \bibinfo{booktitle}{\emph{IEEE
  International Conference on Software Maintenance and Evolution}}.
  \bibinfo{pages}{593--604}.
\newblock
\urldef\tempurl%
\url{https://doi.org/10.1109/ICSME46990.2020.00062}
\showDOI{\tempurl}


\bibitem[Hayes et~al\mbox{.}(2006)]%
        {DBLP:journals/tse/HayesDS06}
\bibfield{author}{\bibinfo{person}{Jane~Huffman Hayes}, \bibinfo{person}{Alex
  Dekhtyar}, {and} \bibinfo{person}{Senthil~Karthikeyan Sundaram}.}
  \bibinfo{year}{2006}\natexlab{}.
\newblock \showarticletitle{Advancing Candidate Link Generation for
  Requirements Tracing: The Study of Methods}.
\newblock \bibinfo{journal}{\emph{{IEEE} Trans. Software Eng.}}
  \bibinfo{volume}{32}, \bibinfo{number}{1} (\bibinfo{year}{2006}),
  \bibinfo{pages}{4--19}.
\newblock


\bibitem[Hey et~al\mbox{.}(2021)]%
        {Hey-icsme2021}
\bibfield{author}{\bibinfo{person}{Tobias Hey}, \bibinfo{person}{Fei Chen},
  \bibinfo{person}{Sebastian Weigelt}, {and} \bibinfo{person}{Walter~F.
  Tichy}.} \bibinfo{year}{2021}\natexlab{}.
\newblock \showarticletitle{Improving Traceability Link Recovery Using
  Fine-grained Requirements-to-Code Relations}. In
  \bibinfo{booktitle}{\emph{{IEEE} International Conference on Software
  Maintenance and Evolution, {ICSME} 2021, Luxembourg, September 27 - October
  1, 2021}}. \bibinfo{publisher}{{IEEE}}, \bibinfo{pages}{12--22}.
\newblock
\urldef\tempurl%
\url{https://doi.org/10.1109/ICSME52107.2021.00008}
\showDOI{\tempurl}


\bibitem[Hoffman and Gelman(2014)]%
        {HoffmanG14}
\bibfield{author}{\bibinfo{person}{Matthew~D. Hoffman} {and}
  \bibinfo{person}{Andrew Gelman}.} \bibinfo{year}{2014}\natexlab{}.
\newblock \showarticletitle{The No-U-turn sampler: adaptively setting path
  lengths in Hamiltonian Monte Carlo}.
\newblock \bibinfo{journal}{\emph{J. Mach. Learn. Res.}} \bibinfo{volume}{15},
  \bibinfo{number}{1} (\bibinfo{year}{2014}), \bibinfo{pages}{1593--1623}.
\newblock
\urldef\tempurl%
\url{https://doi.org/10.5555/2627435.2638586}
\showDOI{\tempurl}


\bibitem[H{\o}st and {\O}stvold(2009)]%
        {DBLP:conf/ecoop/HostO09}
\bibfield{author}{\bibinfo{person}{Einar~W. H{\o}st} {and}
  \bibinfo{person}{Bjarte~M. {\O}stvold}.} \bibinfo{year}{2009}\natexlab{}.
\newblock \showarticletitle{Debugging Method Names}. In
  \bibinfo{booktitle}{\emph{23rd European Conference on Object-Oriented
  Programming}} \emph{(\bibinfo{series}{LNCS}, Vol.~\bibinfo{volume}{5653})},
  \bibfield{editor}{\bibinfo{person}{Sophia Drossopoulou}} (Ed.).
  \bibinfo{publisher}{Springer}, \bibinfo{pages}{294--317}.
\newblock
\urldef\tempurl%
\url{https://doi.org/10.1007/978-3-642-03013-0_14}
\showDOI{\tempurl}


\bibitem[Ivkovic and Kontogiannis(2004)]%
        {Ivkovic/ICSM04}
\bibfield{author}{\bibinfo{person}{I. Ivkovic} {and} \bibinfo{person}{K.
  Kontogiannis}.} \bibinfo{year}{2004}\natexlab{}.
\newblock \showarticletitle{Tracing evolution changes of software artifacts
  through model synchronization}. In \bibinfo{booktitle}{\emph{20th IEEE
  International Conference on Software Maintenance}}.
  \bibinfo{pages}{252--261}.
\newblock
\urldef\tempurl%
\url{https://doi.org/10.1109/ICSM.2004.1357809}
\showDOI{\tempurl}


\bibitem[Kuang et~al\mbox{.}(2019)]%
        {DBLP:conf/iwpc/KuangG0M0ME19}
\bibfield{author}{\bibinfo{person}{Hongyu Kuang}, \bibinfo{person}{Hui Gao},
  \bibinfo{person}{Hao Hu}, \bibinfo{person}{Xiaoxing Ma},
  \bibinfo{person}{Jian Lu}, \bibinfo{person}{Patrick M{\"{a}}der}, {and}
  \bibinfo{person}{Alexander Egyed}.} \bibinfo{year}{2019}\natexlab{}.
\newblock \showarticletitle{Using frugal user feedback with closeness analysis
  on code to improve IR-based traceability recovery}. In
  \bibinfo{booktitle}{\emph{27th International Conference on Program
  Comprehension}}, \bibfield{editor}{\bibinfo{person}{Yann{-}Ga{\"{e}}l
  Gu{\'{e}}h{\'{e}}neuc}, \bibinfo{person}{Foutse Khomh}, {and}
  \bibinfo{person}{Federica Sarro}} (Eds.). \bibinfo{publisher}{{IEEE}/{ACM}},
  \bibinfo{pages}{369--379}.
\newblock


\bibitem[Kuang et~al\mbox{.}(2017)]%
        {DBLP:conf/wcre/KuangNHRLEM17}
\bibfield{author}{\bibinfo{person}{Hongyu Kuang}, \bibinfo{person}{Jia Nie},
  \bibinfo{person}{Hao Hu}, \bibinfo{person}{Patrick Rempel},
  \bibinfo{person}{Jian Lu}, \bibinfo{person}{Alexander Egyed}, {and}
  \bibinfo{person}{Patrick M{\"{a}}der}.} \bibinfo{year}{2017}\natexlab{}.
\newblock \showarticletitle{Analyzing closeness of code dependencies for
  improving IR-based Traceability Recovery}. In \bibinfo{booktitle}{\emph{24th
  {IEEE} International Conference on Software Analysis, Evolution and
  Reengineering}}, \bibfield{editor}{\bibinfo{person}{Martin Pinzger},
  \bibinfo{person}{Gabriele Bavota}, {and} \bibinfo{person}{Andrian Marcus}}
  (Eds.). \bibinfo{publisher}{{IEEE}}, \bibinfo{pages}{68--78}.
\newblock


\bibitem[Le et~al\mbox{.}(2015)]%
        {DBLP:conf/iwpc/LeVLP15}
\bibfield{author}{\bibinfo{person}{Tien{-}Duy~B. Le},
  \bibinfo{person}{Mario~Linares V{\'{a}}squez}, \bibinfo{person}{David Lo},
  {and} \bibinfo{person}{Denys Poshyvanyk}.} \bibinfo{year}{2015}\natexlab{}.
\newblock \showarticletitle{RCLinker: automated linking of issue reports and
  commits leveraging rich contextual information}. In
  \bibinfo{booktitle}{\emph{23rd {IEEE} International Conference on Program
  Comprehension}}. \bibinfo{publisher}{{IEEE}}, \bibinfo{pages}{36--47}.
\newblock
\urldef\tempurl%
\url{https://doi.org/10.1109/ICPC.2015.13}
\showDOI{\tempurl}


\bibitem[Lin et~al\mbox{.}(2021a)]%
        {conf/icse/LinLZ0C21}
\bibfield{author}{\bibinfo{person}{Jinfeng Lin}, \bibinfo{person}{Yalin Liu},
  \bibinfo{person}{Qingkai Zeng}, \bibinfo{person}{Meng Jiang}, {and}
  \bibinfo{person}{Jane Cleland{-}Huang}.} \bibinfo{year}{2021}\natexlab{a}.
\newblock \showarticletitle{Traceability Transformed: Generating more Accurate
  Links with Pre-Trained {BERT} Models}. In \bibinfo{booktitle}{\emph{43rd
  {IEEE/ACM} International Conference on Software Engineering}}.
  \bibinfo{publisher}{{IEEE}}, \bibinfo{pages}{324--335}.
\newblock
\urldef\tempurl%
\url{https://doi.org/10.1109/ICSE43902.2021.00040}
\showDOI{\tempurl}


\bibitem[Lin et~al\mbox{.}(2021b)]%
        {DBLP:conf/icse/LinLZ0C21}
\bibfield{author}{\bibinfo{person}{Jinfeng Lin}, \bibinfo{person}{Yalin Liu},
  \bibinfo{person}{Qingkai Zeng}, \bibinfo{person}{Meng Jiang}, {and}
  \bibinfo{person}{Jane Cleland{-}Huang}.} \bibinfo{year}{2021}\natexlab{b}.
\newblock \showarticletitle{Traceability Transformed: Generating more Accurate
  Links with Pre-Trained {BERT} Models}. In \bibinfo{booktitle}{\emph{43rd
  {IEEE/ACM} International Conference on Software Engineering}}.
  \bibinfo{publisher}{{IEEE}}, \bibinfo{pages}{324--335}.
\newblock
\urldef\tempurl%
\url{https://doi.org/10.1109/ICSE43902.2021.00040}
\showDOI{\tempurl}


\bibitem[Lucia et~al\mbox{.}(2012)]%
        {ADeLuciaSST2012}
\bibfield{author}{\bibinfo{person}{Andrea~De Lucia}, \bibinfo{person}{Andrian
  Marcus}, \bibinfo{person}{Rocco Oliveto}, {and} \bibinfo{person}{Denys
  Poshyvanyk}.} \bibinfo{year}{2012}\natexlab{}.
\newblock \showarticletitle{Information Retrieval Methods for Automated
  Traceability Recovery}.
\newblock In \bibinfo{booktitle}{\emph{Software and Systems Traceability}},
  \bibfield{editor}{\bibinfo{person}{Jane Cleland{-}Huang},
  \bibinfo{person}{Olly Gotel}, {and} \bibinfo{person}{Andrea Zisman}} (Eds.).
  \bibinfo{publisher}{Springer}, \bibinfo{pages}{71--98}.
\newblock
\urldef\tempurl%
\url{https://doi.org/10.1007/978-1-4471-2239-5\_4}
\showDOI{\tempurl}


\bibitem[Macbeth et~al\mbox{.}(2011)]%
        {MacbethCliff}
\bibfield{author}{\bibinfo{person}{Guillermo Macbeth}, \bibinfo{person}{Eugenia
  Razumiejczyk}, {and} \bibinfo{person}{Rubén~Daniel Ledesma}.}
  \bibinfo{year}{2011}\natexlab{}.
\newblock \showarticletitle{Cliff's Delta Calculator: A non-parametric effect
  size program for two groups of observations}.
\newblock \bibinfo{journal}{\emph{Universitas Psychologica}}
  \bibinfo{volume}{10} (\bibinfo{year}{2011}), \bibinfo{pages}{545--555}.
\newblock


\bibitem[M{\"{a}}der and Egyed(2015)]%
        {journals/ese/MaderE15}
\bibfield{author}{\bibinfo{person}{Patrick M{\"{a}}der} {and}
  \bibinfo{person}{Alexander Egyed}.} \bibinfo{year}{2015}\natexlab{}.
\newblock \showarticletitle{Do developers benefit from requirements
  traceability when evolving and maintaining a software system?}
\newblock \bibinfo{journal}{\emph{Empir. Softw. Eng.}} \bibinfo{volume}{20},
  \bibinfo{number}{2} (\bibinfo{year}{2015}), \bibinfo{pages}{413--441}.
\newblock
\urldef\tempurl%
\url{https://doi.org/10.1007/s10664-014-9314-z}
\showDOI{\tempurl}


\bibitem[M{\"{a}}der et~al\mbox{.}(2013)]%
        {DBLP:journals/software/MaderJZC13}
\bibfield{author}{\bibinfo{person}{Patrick M{\"{a}}der},
  \bibinfo{person}{Paul~L. Jones}, \bibinfo{person}{Yi Zhang}, {and}
  \bibinfo{person}{Jane Cleland{-}Huang}.} \bibinfo{year}{2013}\natexlab{}.
\newblock \showarticletitle{Strategic Traceability for Safety-Critical
  Projects}.
\newblock \bibinfo{journal}{\emph{{IEEE} Softw.}} \bibinfo{volume}{30},
  \bibinfo{number}{3} (\bibinfo{year}{2013}), \bibinfo{pages}{58--66}.
\newblock
\urldef\tempurl%
\url{https://doi.org/10.1109/MS.2013.60}
\showDOI{\tempurl}


\bibitem[Maletic and Collard(2015)]%
        {7203124}
\bibfield{author}{\bibinfo{person}{Jonathan~I. Maletic} {and}
  \bibinfo{person}{Michael~L. Collard}.} \bibinfo{year}{2015}\natexlab{}.
\newblock \showarticletitle{Exploration, Analysis, and Manipulation of Source
  Code Using srcML}. In \bibinfo{booktitle}{\emph{37th {IEEE/ACM} International
  Conference on Software Engineering}}, Vol.~\bibinfo{volume}{2}.
  \bibinfo{pages}{951--952}.
\newblock
\urldef\tempurl%
\url{https://doi.org/10.1109/ICSE.2015.302}
\showDOI{\tempurl}


\bibitem[Manning et~al\mbox{.}(2014)]%
        {DBLP:conf/acl/ManningSBFBM14}
\bibfield{author}{\bibinfo{person}{Christopher~D. Manning},
  \bibinfo{person}{Mihai Surdeanu}, \bibinfo{person}{John Bauer},
  \bibinfo{person}{Jenny~Rose Finkel}, \bibinfo{person}{Steven Bethard}, {and}
  \bibinfo{person}{David McClosky}.} \bibinfo{year}{2014}\natexlab{}.
\newblock \showarticletitle{The Stanford {CoreNLP} Natural Language Processing
  Toolkit}. In \bibinfo{booktitle}{\emph{52nd Annual Meeting of the Association
  for Computational Linguistics}}. \bibinfo{publisher}{ACL},
  \bibinfo{pages}{55--60}.
\newblock
\urldef\tempurl%
\url{https://doi.org/10.3115/v1/p14-5010}
\showDOI{\tempurl}


\bibitem[Marcus and Maletic(2003)]%
        {DBLP:conf/icse/MarcusM03}
\bibfield{author}{\bibinfo{person}{Andrian Marcus} {and}
  \bibinfo{person}{Jonathan~I. Maletic}.} \bibinfo{year}{2003}\natexlab{}.
\newblock \showarticletitle{Recovering Documentation-to-Source-Code
  Traceability Links using Latent Semantic Indexing}. In
  \bibinfo{booktitle}{\emph{25th International Conference on Software
  Engineering}}, \bibfield{editor}{\bibinfo{person}{Lori~A. Clarke},
  \bibinfo{person}{Laurie Dillon}, {and} \bibinfo{person}{Walter~F. Tichy}}
  (Eds.). \bibinfo{publisher}{{IEEE}}, \bibinfo{pages}{125--137}.
\newblock


\bibitem[Mayr{-}Dorn et~al\mbox{.}(2021)]%
        {conf/icse/Mayr-DornVBKCEM21}
\bibfield{author}{\bibinfo{person}{Christoph Mayr{-}Dorn},
  \bibinfo{person}{Michael Vierhauser}, \bibinfo{person}{Stefan Bichler},
  \bibinfo{person}{Felix Keplinger}, \bibinfo{person}{Jane Cleland{-}Huang},
  \bibinfo{person}{Alexander Egyed}, {and} \bibinfo{person}{Thomas Mehofer}.}
  \bibinfo{year}{2021}\natexlab{}.
\newblock \showarticletitle{Supporting Quality Assurance with Automated
  Process-Centric Quality Constraints Checking}. In
  \bibinfo{booktitle}{\emph{43rd {IEEE/ACM} International Conference on
  Software Engineering}}. \bibinfo{publisher}{{IEEE}},
  \bibinfo{pages}{1298--1310}.
\newblock
\urldef\tempurl%
\url{https://doi.org/10.1109/ICSE43902.2021.00118}
\showDOI{\tempurl}


\bibitem[McMillan et~al\mbox{.}(2009)]%
        {DBLP:conf/icse/McMillanPR09}
\bibfield{author}{\bibinfo{person}{Collin McMillan}, \bibinfo{person}{Denys
  Poshyvanyk}, {and} \bibinfo{person}{Meghan Revelle}.}
  \bibinfo{year}{2009}\natexlab{}.
\newblock \showarticletitle{Combining textual and structural analysis of
  software artifacts for traceability link recovery}. In
  \bibinfo{booktitle}{\emph{{ICSE} Workshop on Traceability in Emerging Forms
  of Software Engineering}}, \bibfield{editor}{\bibinfo{person}{Giuliano
  Antoniol}, \bibinfo{person}{Denys Poshyvanyk}, {and} \bibinfo{person}{Rocco
  Oliveto}} (Eds.). \bibinfo{publisher}{{IEEE}}, \bibinfo{pages}{41--48}.
\newblock


\bibitem[Mills et~al\mbox{.}(2019)]%
        {conf/icsm/MillsEBKCH19}
\bibfield{author}{\bibinfo{person}{Chris Mills}, \bibinfo{person}{Javier
  Escobar{-}Avila}, \bibinfo{person}{Aditya Bhattacharya},
  \bibinfo{person}{Grigoriy Kondyukov}, \bibinfo{person}{Shayok Chakraborty},
  {and} \bibinfo{person}{Sonia Haiduc}.} \bibinfo{year}{2019}\natexlab{}.
\newblock \showarticletitle{Tracing with Less Data: Active Learning for
  Classification-Based Traceability Link Recovery}. In
  \bibinfo{booktitle}{\emph{2019 {IEEE} International Conference on Software
  Maintenance and Evolution}}. \bibinfo{publisher}{{IEEE}},
  \bibinfo{pages}{103--113}.
\newblock
\urldef\tempurl%
\url{https://doi.org/10.1109/ICSME.2019.00020}
\showDOI{\tempurl}


\bibitem[Mills et~al\mbox{.}(2018)]%
        {conf/icsm/MillsEH18}
\bibfield{author}{\bibinfo{person}{Chris Mills}, \bibinfo{person}{Javier
  Escobar{-}Avila}, {and} \bibinfo{person}{Sonia Haiduc}.}
  \bibinfo{year}{2018}\natexlab{}.
\newblock \showarticletitle{Automatic Traceability Maintenance via Machine
  Learning Classification}. In \bibinfo{booktitle}{\emph{2018 {IEEE}
  International Conference on Software Maintenance and Evolution}}.
  \bibinfo{publisher}{{IEEE}}, \bibinfo{pages}{369--380}.
\newblock
\urldef\tempurl%
\url{https://doi.org/10.1109/ICSME.2018.00045}
\showDOI{\tempurl}


\bibitem[Moran et~al\mbox{.}(2020)]%
        {DBLP:conf/icse/MoranPBMPSJ20}
\bibfield{author}{\bibinfo{person}{Kevin Moran}, \bibinfo{person}{David~N.
  Palacio}, \bibinfo{person}{Carlos Bernal{-}C{\'{a}}rdenas},
  \bibinfo{person}{Daniel McCrystal}, \bibinfo{person}{Denys Poshyvanyk},
  \bibinfo{person}{Chris Shenefiel}, {and} \bibinfo{person}{Jeff Johnson}.}
  \bibinfo{year}{2020}\natexlab{}.
\newblock \showarticletitle{Improving the effectiveness of traceability link
  recovery using hierarchical bayesian networks}. In
  \bibinfo{booktitle}{\emph{42nd International Conference on Software
  Engineering}}, \bibfield{editor}{\bibinfo{person}{Gregg Rothermel} {and}
  \bibinfo{person}{Doo{-}Hwan Bae}} (Eds.). \bibinfo{publisher}{{ACM}},
  \bibinfo{pages}{873--885}.
\newblock
\urldef\tempurl%
\url{https://doi.org/10.1145/3377811.3380418}
\showDOI{\tempurl}


\bibitem[Nejati et~al\mbox{.}(2012)]%
        {DBLP:journals/infsof/NejatiSFBC12}
\bibfield{author}{\bibinfo{person}{Shiva Nejati}, \bibinfo{person}{Mehrdad
  Sabetzadeh}, \bibinfo{person}{Davide Falessi}, \bibinfo{person}{Lionel~C.
  Briand}, {and} \bibinfo{person}{Thierry Coq}.}
  \bibinfo{year}{2012}\natexlab{}.
\newblock \showarticletitle{A SysML-based approach to traceability management
  and design slicing in support of safety certification: Framework, tool
  support, and case studies}.
\newblock \bibinfo{journal}{\emph{Inf. Softw. Technol.}} \bibinfo{volume}{54},
  \bibinfo{number}{6} (\bibinfo{year}{2012}), \bibinfo{pages}{569--590}.
\newblock
\urldef\tempurl%
\url{https://doi.org/10.1016/j.infsof.2012.01.005}
\showDOI{\tempurl}


\bibitem[Newman et~al\mbox{.}(2020)]%
        {Newman-jss2020}
\bibfield{author}{\bibinfo{person}{Christian~D. Newman},
  \bibinfo{person}{Reem~S. Alsuhaibani}, \bibinfo{person}{Michael~John Decker},
  \bibinfo{person}{Anthony Peruma}, \bibinfo{person}{Dishant Kaushik},
  \bibinfo{person}{Mohamed~Wiem Mkaouer}, {and} \bibinfo{person}{Emily Hill}.}
  \bibinfo{year}{2020}\natexlab{}.
\newblock \showarticletitle{On the generation, structure, and semantics of
  grammar patterns in source code identifiers}.
\newblock \bibinfo{journal}{\emph{J. Syst. Softw.}}  \bibinfo{volume}{170}
  (\bibinfo{year}{2020}), \bibinfo{pages}{110740}.
\newblock
\urldef\tempurl%
\url{https://doi.org/10.1016/j.jss.2020.110740}
\showDOI{\tempurl}


\bibitem[Nhlabatsi et~al\mbox{.}(2015)]%
        {DBLP:conf/icse/NhlabatsiYZTKBK15}
\bibfield{author}{\bibinfo{person}{Armstrong Nhlabatsi}, \bibinfo{person}{Yijun
  Yu}, \bibinfo{person}{Andrea Zisman}, \bibinfo{person}{Thein~Than Tun},
  \bibinfo{person}{Niamul Khan}, \bibinfo{person}{Arosha~K. Bandara},
  \bibinfo{person}{Khaled~M. Khan}, {and} \bibinfo{person}{Bashar Nuseibeh}.}
  \bibinfo{year}{2015}\natexlab{}.
\newblock \showarticletitle{Managing Security Control Assumptions Using Causal
  Traceability}. In \bibinfo{booktitle}{\emph{8th {IEEE/ACM} International
  Symposium on Software and Systems Traceability}},
  \bibfield{editor}{\bibinfo{person}{Patrick M{\"{a}}der} {and}
  \bibinfo{person}{Rocco Oliveto}} (Eds.). \bibinfo{publisher}{{IEEE} Computer
  Society}, \bibinfo{pages}{43--49}.
\newblock
\urldef\tempurl%
\url{https://doi.org/10.1109/SST.2015.14}
\showDOI{\tempurl}


\bibitem[Nishikawa et~al\mbox{.}(2015)]%
        {conf/icsm/NishikawaWFOM15}
\bibfield{author}{\bibinfo{person}{Kazuki Nishikawa}, \bibinfo{person}{Hironori
  Washizaki}, \bibinfo{person}{Yoshiaki Fukazawa}, \bibinfo{person}{Keishi
  Oshima}, {and} \bibinfo{person}{Ryota Mibe}.}
  \bibinfo{year}{2015}\natexlab{}.
\newblock \showarticletitle{Recovering transitive traceability links among
  software artifacts}. In \bibinfo{booktitle}{\emph{{IEEE} International
  Conference on Software Maintenance and Evolution}}.
  \bibinfo{publisher}{{IEEE}}, \bibinfo{pages}{576--580}.
\newblock
\urldef\tempurl%
\url{https://doi.org/10.1109/ICSM.2015.7332517}
\showDOI{\tempurl}


\bibitem[Panichella et~al\mbox{.}(2013a)]%
        {Panichella6606598}
\bibfield{author}{\bibinfo{person}{Annibale Panichella},
  \bibinfo{person}{Bogdan Dit}, \bibinfo{person}{Rocco Oliveto},
  \bibinfo{person}{Massimilano Di~Penta}, \bibinfo{person}{Denys Poshynanyk},
  {and} \bibinfo{person}{Andrea De~Lucia}.} \bibinfo{year}{2013}\natexlab{a}.
\newblock \showarticletitle{How to effectively use topic models for software
  engineering tasks? An approach based on Genetic Algorithms}. In
  \bibinfo{booktitle}{\emph{35th International Conference on Software
  Engineering}}. \bibinfo{pages}{522--531}.
\newblock
\urldef\tempurl%
\url{https://doi.org/10.1109/ICSE.2013.6606598}
\showDOI{\tempurl}


\bibitem[Panichella et~al\mbox{.}(2015)]%
        {DBLP:conf/icse/PanichellaLZ15}
\bibfield{author}{\bibinfo{person}{Annibale Panichella},
  \bibinfo{person}{Andrea~De Lucia}, {and} \bibinfo{person}{Andy Zaidman}.}
  \bibinfo{year}{2015}\natexlab{}.
\newblock \showarticletitle{Adaptive User Feedback for IR-Based Traceability
  Recovery}. In \bibinfo{booktitle}{\emph{8th {IEEE/ACM} International
  Symposium on Software and Systems Traceability}},
  \bibfield{editor}{\bibinfo{person}{Patrick M{\"{a}}der} {and}
  \bibinfo{person}{Rocco Oliveto}} (Eds.). \bibinfo{publisher}{{IEEE}},
  \bibinfo{pages}{15--21}.
\newblock


\bibitem[Panichella et~al\mbox{.}(2013b)]%
        {DBLP:conf/csmr/PanichellaMMPOPL13}
\bibfield{author}{\bibinfo{person}{Annibale Panichella},
  \bibinfo{person}{Collin McMillan}, \bibinfo{person}{Evan Moritz},
  \bibinfo{person}{Davide Palmieri}, \bibinfo{person}{Rocco Oliveto},
  \bibinfo{person}{Denys Poshyvanyk}, {and} \bibinfo{person}{Andrea~De Lucia}.}
  \bibinfo{year}{2013}\natexlab{b}.
\newblock \showarticletitle{When and How Using Structural Information to
  Improve IR-Based Traceability Recovery}. In \bibinfo{booktitle}{\emph{17th
  European Conference on Software Maintenance and Reengineering}},
  \bibfield{editor}{\bibinfo{person}{Anthony Cleve}, \bibinfo{person}{Filippo
  Ricca}, {and} \bibinfo{person}{Maura Cerioli}} (Eds.).
  \bibinfo{publisher}{{IEEE}}, \bibinfo{pages}{199--208}.
\newblock


\bibitem[Porter(1980)]%
        {DBLP:journals/program/Porter80}
\bibfield{author}{\bibinfo{person}{Martin~F. Porter}.}
  \bibinfo{year}{1980}\natexlab{}.
\newblock \showarticletitle{An algorithm for suffix stripping}.
\newblock \bibinfo{journal}{\emph{Program}} \bibinfo{volume}{14},
  \bibinfo{number}{3} (\bibinfo{year}{1980}), \bibinfo{pages}{130--137}.
\newblock
\urldef\tempurl%
\url{https://doi.org/10.1108/eb046814}
\showDOI{\tempurl}


\bibitem[Poshyvanyk et~al\mbox{.}(2007)]%
        {DBLP:journals/tse/PoshyvanykGMAR07}
\bibfield{author}{\bibinfo{person}{Denys Poshyvanyk},
  \bibinfo{person}{Yann{-}Ga{\"{e}}l Gu{\'{e}}h{\'{e}}neuc},
  \bibinfo{person}{Andrian Marcus}, \bibinfo{person}{Giuliano Antoniol}, {and}
  \bibinfo{person}{V{\'{a}}clav Rajlich}.} \bibinfo{year}{2007}\natexlab{}.
\newblock \showarticletitle{Feature Location Using Probabilistic Ranking of
  Methods Based on Execution Scenarios and Information Retrieval}.
\newblock \bibinfo{journal}{\emph{{IEEE} Trans. Software Eng.}}
  \bibinfo{volume}{33}, \bibinfo{number}{6} (\bibinfo{year}{2007}),
  \bibinfo{pages}{420--432}.
\newblock


\bibitem[Ra\c{t}iu et~al\mbox{.}(2022)]%
        {Cosmina2022}
\bibfield{author}{\bibinfo{person}{Cosmina~Cristina Ra\c{t}iu},
  \bibinfo{person}{Wesley K.~G. Assun\c{c}\~{a}o}, \bibinfo{person}{Rainer
  Haas}, {and} \bibinfo{person}{Alexander Egyed}.}
  \bibinfo{year}{2022}\natexlab{}.
\newblock \showarticletitle{Reactive Links across Multi-Domain Engineering
  Models}. In \bibinfo{booktitle}{\emph{25th International Conference on Model
  Driven Engineering Languages and Systems}}. \bibinfo{publisher}{ACM},
  \bibinfo{pages}{76–86}.
\newblock
\showISBNx{9781450394666}
\urldef\tempurl%
\url{https://doi.org/10.1145/3550355.3552446}
\showDOI{\tempurl}


\bibitem[Ramesh and Jarke(2001)]%
        {DBLP:journals/tse/RameshJ01}
\bibfield{author}{\bibinfo{person}{Balasubramaniam Ramesh} {and}
  \bibinfo{person}{Matthias Jarke}.} \bibinfo{year}{2001}\natexlab{}.
\newblock \showarticletitle{Toward Reference Models of Requirements
  Traceability}.
\newblock \bibinfo{journal}{\emph{{IEEE} Trans. Software Eng.}}
  \bibinfo{volume}{27}, \bibinfo{number}{1} (\bibinfo{year}{2001}),
  \bibinfo{pages}{58--93}.
\newblock
\urldef\tempurl%
\url{https://doi.org/10.1109/32.895989}
\showDOI{\tempurl}


\bibitem[Rath et~al\mbox{.}(2018)]%
        {conf/msr/0002LM08}
\bibfield{author}{\bibinfo{person}{Michael Rath}, \bibinfo{person}{David Lo},
  {and} \bibinfo{person}{Patrick M{\"{a}}der}.}
  \bibinfo{year}{2018}\natexlab{}.
\newblock \showarticletitle{Analyzing requirements and traceability information
  to improve bug localization}. In \bibinfo{booktitle}{\emph{15th International
  Conference on Mining Software Repositories}},
  \bibfield{editor}{\bibinfo{person}{Andy Zaidman}, \bibinfo{person}{Yasutaka
  Kamei}, {and} \bibinfo{person}{Emily Hill}} (Eds.).
  \bibinfo{publisher}{{ACM}}, \bibinfo{pages}{442--453}.
\newblock
\urldef\tempurl%
\url{https://doi.org/10.1145/3196398.3196415}
\showDOI{\tempurl}


\bibitem[Rath et~al\mbox{.}(2019)]%
        {DBLP:conf/se/0002RGCM19}
\bibfield{author}{\bibinfo{person}{Michael Rath}, \bibinfo{person}{Jacob
  Rendall}, \bibinfo{person}{Jin L.~C. Guo}, \bibinfo{person}{Jane
  Cleland{-}Huang}, {and} \bibinfo{person}{Patrick M{\"{a}}der}.}
  \bibinfo{year}{2019}\natexlab{}.
\newblock \showarticletitle{Traceability in the Wild: Automatically Augmenting
  Incomplete Trace Links}. In \bibinfo{booktitle}{\emph{Software Engineering
  and Software Management}} \emph{(\bibinfo{series}{{LNI}},
  Vol.~\bibinfo{volume}{{P-292}})}. \bibinfo{publisher}{{GI}},
  \bibinfo{pages}{63}.
\newblock


\bibitem[Rempel and M{\"{a}}der(2017)]%
        {DBLP:journals/tse/RempelM17}
\bibfield{author}{\bibinfo{person}{Patrick Rempel} {and}
  \bibinfo{person}{Patrick M{\"{a}}der}.} \bibinfo{year}{2017}\natexlab{}.
\newblock \showarticletitle{Preventing Defects: The Impact of Requirements
  Traceability Completeness on Software Quality}.
\newblock \bibinfo{journal}{\emph{{IEEE} Trans. Software Eng.}}
  \bibinfo{volume}{43}, \bibinfo{number}{8} (\bibinfo{year}{2017}),
  \bibinfo{pages}{777--797}.
\newblock


\bibitem[Rodriguez et~al\mbox{.}(2021)]%
        {conf/icsm/RodriguezCF21}
\bibfield{author}{\bibinfo{person}{Alberto~D. Rodriguez}, \bibinfo{person}{Jane
  Cleland{-}Huang}, {and} \bibinfo{person}{Davide Falessi}.}
  \bibinfo{year}{2021}\natexlab{}.
\newblock \showarticletitle{Leveraging Intermediate Artifacts to Improve
  Automated Trace Link Retrieval}. In \bibinfo{booktitle}{\emph{{IEEE}
  International Conference on Software Maintenance and Evolution}}.
  \bibinfo{publisher}{{IEEE}}, \bibinfo{pages}{81--92}.
\newblock
\urldef\tempurl%
\url{https://doi.org/10.1109/ICSME52107.2021.00014}
\showDOI{\tempurl}


\bibitem[Srikanth and Srihari(2002)]%
        {10.1145/564376.564476BTIR}
\bibfield{author}{\bibinfo{person}{Munirathnam Srikanth} {and}
  \bibinfo{person}{Rohini Srihari}.} \bibinfo{year}{2002}\natexlab{}.
\newblock \showarticletitle{Biterm Language Models for Document Retrieval}. In
  \bibinfo{booktitle}{\emph{25th Annual International ACM SIGIR Conference on
  Research and Development in Information Retrieval}}.
  \bibinfo{publisher}{ACM}, \bibinfo{pages}{425–426}.
\newblock
\showISBNx{1581135610}
\urldef\tempurl%
\url{https://doi.org/10.1145/564376.564476}
\showDOI{\tempurl}


\bibitem[Sun et~al\mbox{.}(2017)]%
        {DBLP:journals/infsof/SunWY17}
\bibfield{author}{\bibinfo{person}{Yan Sun}, \bibinfo{person}{Qing Wang}, {and}
  \bibinfo{person}{Ye Yang}.} \bibinfo{year}{2017}\natexlab{}.
\newblock \showarticletitle{FRLink: Improving the recovery of missing
  issue-commit links by revisiting file relevance}.
\newblock \bibinfo{journal}{\emph{Inf. Softw. Technol.}}  \bibinfo{volume}{84}
  (\bibinfo{year}{2017}), \bibinfo{pages}{33--47}.
\newblock
\urldef\tempurl%
\url{https://doi.org/10.1016/j.infsof.2016.11.010}
\showDOI{\tempurl}


\bibitem[Wilcoxon(1944)]%
        {WilcoxonIndividual}
\bibfield{author}{\bibinfo{person}{Frank Wilcoxon}.}
  \bibinfo{year}{1944}\natexlab{}.
\newblock \showarticletitle{Individual Comparisons by Ranking Methods. Biom
  Bull}.
\newblock \bibinfo{journal}{\emph{Biometrics}} \bibinfo{volume}{1},
  \bibinfo{number}{6} (\bibinfo{year}{1944}), \bibinfo{pages}{80--83}.
\newblock


\end{thebibliography}
\end{document}